\documentclass[traditabstract]{aa} 

\usepackage{graphicx}
\usepackage{txfonts}
\usepackage{amssymb}
\usepackage{array}
\usepackage{dcolumn}
\usepackage{natbib}
\usepackage{multirow}
\usepackage{txfonts}
\usepackage{natbib}
\bibpunct{(}{)}{;}{a}{}{,}
\begin{document} 
\title{CO map and steep Kennicutt-Schmidt relation in the extended UV disk of 
M63
\thanks{Based on observations carried out with the IRAM 30~m telescope. IRAM 
is supported by INSU/CNRS (France), MPG (Germany), and IGN (Spain).}}


\author{M. Dessauges-Zavadsky\inst{1},
        C. Verdugo\inst{2},
        F. Combes\inst{2},
        \and
        D. Pfenniger\inst{1}
        }

\institute{Observatoire de Gen\`eve, Universit\'e de Gen\`eve, 
           51, Ch. des Maillettes, 1290 Sauverny, Switzerland\\
           \email{miroslava.dessauges@unige.ch, daniel.pfenniger@unige.ch}
         \and
           Observatoire de Paris, LERMA, 61 Av. de l'Observatoire, 
           75014 Paris, France\\
           \email{celia.verdugo@obspm.fr, francoise.combes@obspm.fr}
           }


\authorrunning{Dessauges-Zavadsky et~al.}

\titlerunning{CO and Kennicutt-Schmidt relation in the XUV disk of M63}


\abstract{
Results from the UV satellite GALEX revealed surprisingly large extensions 
of disks in some nearby spiral galaxies. While the H$\alpha$ emission, the 
usual tracer of star formation, drops down at the border of the isophotal 
radius, $r_{25}$, the UV emission extends out to 3 to 4 times this radius and 
often covers a significant fraction of the H\,I area. M63 is a remarkable 
example of a spiral galaxy with one of the most extended UV disks, so it offers 
the opportunity to search for the molecular gas and characterize the star 
formation in outer disk regions as revealed by the UV emission. We obtained 
deep CO(1--0) and CO(2--1) observations on the IRAM 30~m telescope along the 
major axis of the M63 disk from the center out to the galactocentric radius 
$r_{\rm gal} = 1.6\,r_{25}$ and over a bright UV region at $r_{\rm gal} = 
1.36\,r_{25}$. CO(1--0) is detected all along the M63 major axis out to 
$r_{25}$, and CO(2--1) is confined to $r_{\rm gal} = 0.68\,r_{25}$, which may 
betray lower excitation temperatures in the outer disk. CO(1--0) is also 
detected in the external bright UV region of M63. This is the fourth molecular 
gas detection in the outskirts of nearby spirals. The radial profiles of the CO 
emission and of the H$\alpha$, 24~$\mu$m, NUV and FUV star formation tracers 
and H\,I taken from the literature show a severe drop with the galactocentric 
radius, such that beyond $r_{25}$ they are all absent with the exception of a 
faint UV emission and H\,I. The CO emission detection in the external UV 
region, where the UV flux is higher than the UV flux observed beyond $r_{25}$, 
highlights a tight correlation between the CO and UV fluxes, namely the amount 
of molecular gas and the intensity of star formation. This external UV region 
is dominated by the atomic gas, suggesting that H\,I is more likely the 
precursor of H$_2$ rather than the product of UV photodissociation. A broken 
power law needs to be invoked to describe the Kennicutt-Schmidt (K-S) relation 
of M63 from the center of the galaxy out to $r_{\rm gal} = 1.36\,r_{25}$. While 
all along the major axis out to $r_{25}$ the K-S relation is almost linear 
(with a slope of nearly 1 in log space), in the external UV region the SFR 
regime is highly nonlinear and characterized by a steep K-S relation (with a 
slope much higher than 1 in log space) and very low star formation efficiency.}

\keywords{Galaxies: star formation -- Ultraviolet: galaxies -- Galaxies: ISM -- Submillimeter: ISM -- Galaxies: evolution}

\maketitle

%

\section{Introduction}

The study of star formation in the outer regions of disks of normal spiral 
galaxies has gained interest in the past few years, mainly because they are 
low-metallicity environments \citep{henry1999}, resembling the conditions of 
early stages of spiral galaxies and high-redshift galaxies. These regions are 
also known to have low star formation rates \citep{dong2008,bigiel2010,
alberts2011}. Moreover, there is growing evidence of cold gas accretion in the 
local Universe, both through the arrival and merging of gas-rich satellites 
and through gas infall from the intergalactic medium. This new gas could be 
deposited in the outer regions of galaxies and form reservoirs for replenishing 
the inner parts and feeding star formation \citep{sancisi2008}, making outer 
regions good laboratories for scanning the interface between galaxies and the 
surrounding intergalactic gas.

With its 1.25 degree field of view and sensitivity to stellar populations 
younger than a few hundred Myr, the UV satellite GALEX (Galaxy Evolution 
Explorer) is well suited to address the question of star formation in spiral 
galaxies at large galactocentric radii. Recent star formation within such 
environments was detected in H$\alpha$ (the principal star formation tracer 
over the years) and broad-band observations for a few galaxies: NGC\,628, 
NGC\,1058, NGC\,6946 \citep{ferguson1998}, M31 \citep{cuillandre2001}, and 
NGC\,6822 \citep{deblok2003}. However, the GALEX far-UV (FUV) and near-UV (NUV) 
data demonstrate that H$\alpha$ observations still fail to detect a significant 
population of moderate-age stars in the outermost disks of spiral galaxies, 
since H$\alpha$ traces more recent star formation episodes. Indeed, UV-bright 
disks extending up to 3 to 4 times the optical radius have been reported in 
about 30\% of spiral galaxies, with the most remarkable examples: M63, M83, 
NGC\,2841, and NGC\,4625 \citep{thilker2005,gildepaz2005,gildepaz2007}. These 
extended UV emission (XUV) disks cover a significant fraction of the area 
detected at 21~cm wavelength, with some correspondence between the position of 
the brightest UV complexes and peaks in the atomic gas distribution. The 
measured ${\rm FUV} - {\rm NUV}$ colors are generally consistent with young 
populations of O, and predominantly B stars, characterized by ages from a few 
Myr up to 400~Myr.

The confirmed occurrence of recent and ongoing star formation in the outer 
disks of normal spiral galaxies has several important implications. First, it 
supports the presence of molecular gas in the outskirts of spirals, since stars 
are formed within molecular clouds. Second, this suggests the presence of large 
reservoirs of hydrogen in the form of H$_2$, which may contribute to the 
baryonic dark matter of spiral galaxies. Third, it offers the ideal place to 
study the unresolved issue of the atomic hydrogen gas origin: is H\,I mainly
a product of the star formation process; i.e., does it result from the 
photodissociation of H$_2$ by the UV flux radiation emanated from newly formed 
stars \citep{allen1986,allen2004,smith2000}, rather than mainly being a 
precursor to it? Fourth, the presence of recently formed stellar complexes at 
large galactocentric radii also provides a simplified laboratory for 
determining the star formation threshold, namely the minimum gas surface 
density required for star formation to occur spontaneously 
\citep{kennicutt1989,martin2001}. Fifth, it allows investigating the star 
formation in quiescent and low-metallicity environments that may affect the 
star formation density and the initial mass function.

In this paper we aim to detect the molecular gas expected in the outskirts of 
spiral galaxies because of the star formation discovered from the XUV 
observations. The M63 spiral galaxy, known to have an XUV disk, is selected for 
this work. The search for molecular gas in galaxies is difficult, since we 
cannot detect cold H$_2$ directly. Instead, the second most common molecule, 
CO, is used as a proxy. Molecular gas has been detected in many galaxies from 
the mapping of the CO emission. It is now well established that the CO emission 
is the strongest in the central regions of spiral galaxies \citep[e.g.,][]
{young1991}, but then falls off, as does the blue stellar light, with the 
galactocentric radius \citep[e.g.,][]{young1982,young1995}. The questions ``is 
this CO emission drop real, or does it reflect the difficulty of detecting the 
molecular gas in the outer regions of spiral galaxies'' remain open. For 
example, CO emission lines are known to be weaker in media characterized by 
low metallicities, low gas temperatures, low excitations, and low gas 
densities, even if substantial molecular gas is present \citep[e.g.,][]
{allen1996,combes2000}, and thus in media that may well be potentially 
representative of outer disk regions.

\citet{schruba2011} used a very effective stacking technique on HERACLES 
CO(2--1) data of nearby spiral galaxies, where they stack CO spectra 
across many sightlines by assuming that the mean H\,I and CO velocities are 
similar, to highlight the presence of the faintest CO emission. They focused 
their analysis on data stacked in bins of galactocentric radius and found that 
the CO radial profile follows a remarkably uniform exponential decline with a 
scale length of $\sim 0.2\,r_{25}$. But even though this has been shown to be 
a very effective technique, it still fails to detect CO in the farthest out 
regions of galactic disks, where they only reach $3\,\sigma$ upper limits. 
Specific very deep CO observations are thus required to trace the molecular 
gas in the outer disk regions of spiral galaxies, which is precisely the 
objective of this work.

%

\begin{figure}
\includegraphics[width=9cm,clip]{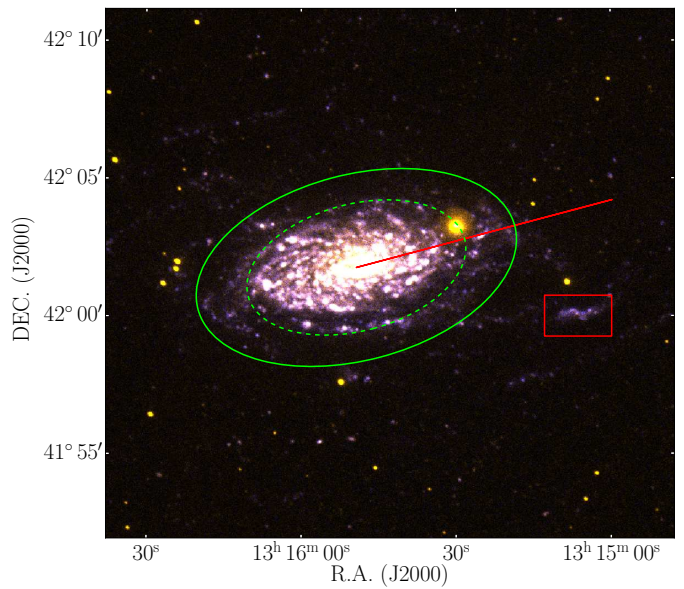}
\caption{False three-color (red/green/blue) composite image of M63 from 
\citet{gildepaz2007}. The image was produced using the {\em arcsinh} function, 
which allows showing faint structures while simultaneously preserving brighter 
structures in the field, such as the spiral arms of large galaxies 
\citep{lupton2004}. 
The green ellipse represents the $B$-band $25~\rm mag~arcsec^{-2}$ isophote 
limit, $r_{25}$. The red solid line shows the CO mapping we performed along the 
major axis of M63 from the center out to $r_{\rm gal} = 572\arcsec = 
1.6\,r_{25}$, where 27 single pointings were aligned with a spatial sampling of 
$22\arcsec$. The red box encloses the bright UV region at $r_{\rm gal} = 
483\arcsec = 1.36\,r_{25}$ which we mapped with $6\times 2 = 12$ pointings with 
the same spatial sampling. The CO(1--0) emission is observed out to the 
isophotal radius along the radial cut and in the bright UV region, while the 
CO(2--1) emission is confined to $0.68\,r_{25}$ (dashed green ellipse).}
\label{fig:M63image}
\end{figure}
%

In Sect.~\ref{sect:M63} we present the characteristics of the M63 spiral 
galaxy, selected for deep CO observations in the outer disk regions. We qualify 
the advantages of M63 in the context of XUV galaxies and cite the previous 
works done on M63. In Sect.~\ref{sect:observations} we describe the CO 
observations performed on the IRAM 30~m telescope and show the corresponding 
results in Sect.~\ref{sect:results}. In Sect.~\ref{sect:discussion} we discuss 
the radial profiles obtained for the acquired CO measurements in comparison to 
other star formation tracers (FUV, NUV, H$\alpha$, 24~$\mu$m) and H\,I, and 
investigate the Kennicutt-Schmidt relations across the galaxy and beyond the 
optical radius. Summary and conclusions are given in 
Sect.~\ref{sect:conclusion}. Throughout the paper we adopt the ``standard'' 
Galactic CO(1--0)-to-H$_2$ conversion factor $X_{\rm CO} = 2\times 
10^{20}~\rm cm^{-2}(K~km~s^{-1})^{-1}$ \citep{dickman1986} and include a 
correction for helium. 

%

\begin{table*}[!]
\caption{CO mapping along the major axis of the M63 disk}             
\label{tab:M63disk}      
\centering          
\begin{tabular}{r l r | D{.}{}{1} D{.}{}{2} c | D{.}{}{1} D{.}{}{2} c | c c c}     
\hline\hline 
 & & & \multicolumn{3}{c|}{CO(1--0)} & \multicolumn{3}{c|}{CO(2--1)} & & & \\
$r_{\rm gal}$ & $r_{\rm gal}$ & Time & \multicolumn{1}{c}{rms\tablefootmark{a}} & \multicolumn{1}{c}{FWHM\tablefootmark{b}} & $F$\tablefootmark{c} & \multicolumn{1}{c}{rms\tablefootmark{a}} & \multicolumn{1}{c}{FWHM\tablefootmark{b}} & $F$\tablefootmark{c} &$r_{2,1}$\tablefootmark{d} & $M_{\rm H_2}$\tablefootmark{e} & $\Sigma_{\rm H_2}$\tablefootmark{f} \\
($\arcsec$) & ($r_{25}$) & (min) & \multicolumn{1}{c}{(mK)} & \multicolumn{1}{c}{(km~s$^{-1}$)} & (K~km~s$^{-1}$) & \multicolumn{1}{c}{(mK)}                & \multicolumn{1}{c}{(km~s$^{-1}$)} & (K~km~s$^{-1}$) &  & ($\rm 10^6~M_{\sun}$) & ($\rm M_{\sun}~pc^2$) \\
\hline 
  0 & 0    &  31 & 11.	     & 200 & $49.8\pm 2.2$  & 16.	& 192 & $39.8\pm 3.1$	    & 0.24    & $321.8\pm 14.2$ & $125.7\pm 5.6$ \\
 22 & 0.06 &   8 & 18.	     &  53 & $24.5\pm 0.9$  & 28.	&  27 & $12.6\pm 0.8$	    & 0.16    & $158.3\pm 5.8$  &  $61.8\pm 2.3$ \\
 44 & 0.12 &   8 & 19.	     &  41 & $16.4\pm 0.8$  & 33.	&  21 & $9.0\pm 0.7$	    & 0.17    & $106.0\pm 5.2$  &  $41.4\pm 2.0$ \\
 66 & 0.19 &   8 & 19.	     &  23 &  $9.8\pm 0.4$  & 32.	&  16 & $4.2\pm 0.5$	    & 0.13    &  $63.3\pm 2.6$  &  $24.7\pm 1.0$ \\
 88 & 0.25 &   8 & 18.	     &  20 &  $6.3\pm 0.4$  & 32.	&  20 & $3.4\pm 0.6$	    & 0.17    &  $40.7\pm 2.6$  &  $15.9\pm 1.0$ \\
110 & 0.31 &   8 & 19.	     &  26 & $12.6\pm 0.5$  & 30.	&  21 & $8.2\pm 0.6$	    & 0.20    &  $81.4\pm 3.2$  &  $31.8\pm 1.3$ \\
132 & 0.37 &   8 & 21.	     &  22 &  $5.1\pm 0.5$  & 35.	&  20 & $2.6\pm 0.7$	    & 0.16    &  $33.0\pm 3.2$  &  $12.9\pm 1.3$ \\
154 & 0.44 &  23 & 13.	     &  29 &  $5.3\pm 0.4$  & 22.	&  39 & $2.5\pm 0.9\ddag$   & 0.15    &  $34.3\pm 2.6$  &  $13.4\pm 1.0$ \\
176 & 0.50 &  23 & 13.	     &  19 &  $5.4\pm 0.2$  & 24.	&  27 & $2.8\pm 0.6$	    & 0.16    &  $34.9\pm 1.3$  &  $13.6\pm 0.5$ \\
198 & 0.56 &  23 & 13.	     &  21 &  $1.8\pm 0.3$  & 22.	&  18 & $0.82\pm 0.40\ddag$ & 0.14    &  $11.6\pm 1.9$  &   $4.5\pm 0.8$ \\
220 & 0.62 &  31 & 11.	     &  16 & $0.98\pm 0.17$ & 11.$\dag$ &  10 & $<0.33$ 	    & $<0.11$ &   $6.3\pm 1.1$  &   $2.5\pm 0.4$ \\
242 & 0.68 & 101 &  6.	     &  11 & $0.30\pm 0.06$ &  8.	&   6 & $0.17\pm 0.05$      & 0.17    &   $1.9\pm 0.4$  &  $0.76\pm 0.15$ \\
264 & 0.75 & 101 &  6.	     &  14 & $0.45\pm 0.08$ &  4.$\dag$ &  10 & $<0.12$ 	    & $<0.08$ &   $2.9\pm 0.5$  &   $1.1\pm 0.2$ \\
286 & 0.81 &  55 & 10.	     &  21 & $0.88\pm 0.21$ &  9.$\dag$ &  10 & $<0.27$ 	    & $<0.10$ &   $5.7\pm 1.4$  &   $2.2\pm 0.5$ \\
308 & 0.87 &  47 & 11.	     &  14 & $0.81\pm 0.15$ & 11.$\dag$ &  10 & $<0.33$ 	    & $<0.13$ &   $5.2\pm 1.0$  &   $2.0\pm 0.4$ \\
330 & 0.93 &  63 &  7.	     &  22 & $0.85\pm 0.15$ &  9.	&  34 & $0.38\pm 0.31\ddag$ & 0.14    &   $5.5\pm 1.0$  &   $2.1\pm 0.4$ \\
352 & 0.99 & 155 &  5.	     &	 9 & $0.23\pm 0.04$ &  6.	&  34 & $0.29\pm 0.21\ddag$ & 0.38    &   $1.5\pm 0.3$  &  $0.58\pm 0.10$ \\
374 & 1.06 & 109 &  4.$\dag$ &  10 & $<0.12$	    &  4.$\dag$ &  10 & $<0.12$ 	    &	      & & \\
396 & 1.12 & 117 &  4.$\dag$ &  10 & $<0.12$	    &  3.$\dag$ &  10 & $<0.09$ 	    &	      & & \\
418 & 1.18 &  93 &  5.$\dag$ &  10 & $<0.15$	    &  5.$\dag$ &  10 & $<0.15$ 	    &	      & & \\
... & ...  &     &           &     &                &           &     &                     &         & & \\
\hline
473 & 1.34 & 381 &  2.$\dag$ &  10 & $<0.06$	    &  2.$\dag$ &  10 & $<0.06$ 	    &	      & $<0.39$         & $<0.15$ \\
\hline                                
\end{tabular}
\tablefoot{
The radius of the $B$-band $25~\rm mag~arcsec^{-2}$ isophote, the so-called 
isophotal radius, is equal to $r_{25} = 354\arcsec = 17.4$~kpc in M63. It is 
often used as a reference to express the relative galactocentric radius, 
$r_{\rm gal}$. The last separate line of the Table gives the values obtained 
when summing all scans at the 10 outermost pointings from $1.06\,r_{25}$ to 
$1.6\,r_{25}$ with no CO detection and by smoothing the resulting 3~mm and 
1~mm spectra to a resolution of $9.7~\rm km~s^{-1}$ and $10.4~\rm km~s^{-1}$, 
respectively.
\tablefoottext{a}{Rms noises at, respectively, 3~mm and 1~mm, measured per 
channel of $3.2~\rm km~s^{-1}$ and $1.3~\rm km~s^{-1}$. The values marked with 
$\dag$ correspond to rms noises measured per smoothed channel of 
$9.7~\rm km~s^{-1}$ and $10.4~\rm km~s^{-1}$, respectively.} 
\tablefoottext{b}{Full widths at half maximum of, respectively, CO(1--0) and 
CO(2--1) lines as determined from fitting Gaussian profiles.} 
\tablefoottext{c}{Integrated CO(1--0) and CO(2-1) line fluxes as determined 
from fitting Gaussian profiles. The values marked with $\ddag$ correspond to 
potentially less reliable measurements, because of a CO line detection at 
$2\,\sigma$ only. Upper limits are $3\,\sigma$ and are calculated assuming a 
$10~\rm km~s^{-1}$ line width.}
\tablefoottext{d}{CO luminosity ratios defined as $r_{2,1} = 
L'_{\rm CO(2-1)}/L'_{\rm CO(1-0)}$, needed to correct the lower Rayleigh-Jeans
brightness temperature of the 2--1 transition relative to 1--0. We would like, 
however, to stress that CO(1--0) and CO(2--1) do not exactly map the same 
regions, because the CO(1--0) beam area equals $4\times$ the CO(2--1) beam 
area. The CO luminosity is calculated using the formula (3) from 
\citet{solomon1997}.}
\tablefoottext{e}{H$_2$ masses calculated from the CO(1--0) luminosity and by 
adopting the ``standard'' Galactic CO-to-H$_2$ conversion factor, $X_{\rm CO} = 
2\times 10^{20}~\rm cm^{-2}~(K~km~s^{-1})^{-1}$ \citep{dickman1986}. The 
applied formula is: 
$M_{\rm H_2}~({\rm M_{\sun}}) = 4.4 L'_{\rm CO(1-0)}~({\rm K~km~s^{-1}~pc^2})$, 
where a factor of 1.36 is included to account for helium.}
\tablefoottext{f}{H$_2$ surface densities calculated from the integrated 
CO(1--0) line flux and by adopting the CO-to-H$_2$ conversion factor given 
under $^{(e)}$. The applied formula is: 
$\Sigma_{\rm H_2}~({\rm M_{\sun}~pc^{-2}}) = 
4.4 \cos(i) F_{\rm CO(1-0)}~({\rm K~km~s^{-1}})$, where $i = 55\degr$ is 
the inclination of M63 \citep{leroy2008}.}
}
\end{table*}
%

\section{M63 characteristics}
\label{sect:M63}

Classified as SA(rs)bc and located at 10.1~Mpc \citep{leroy2009}, M63 (or 
NGC5055) looks like a typical spiral galaxy, representative of a large class of 
local spirals, with no immediate neighbour, which excludes the potentiality of 
a galaxy in interaction. However, M63 is not very ordinary, since it is a 
remarkable example of a nearby spiral with a bright and XUV disk. In 
Fig.~\ref{fig:M63image} we show the GALEX NUV and FUV color-composite image of 
M63, where an extensive population of UV-bright star-forming regions and 
stellar clusters (tracing the O and B stars) is revealed. The M63 NUV and FUV
surface-brightness profiles show a smooth decrease in the UV emission out to 
$700\arcsec$ in the equivalent radius (defined as the square root of the 
product of the half-minor axis times the half-major axis), namely out to 2.5 
times the optical equivalent radius \citep{gildepaz2007}. The major axis radius 
of the $B$-band $25~\rm mag~arcsec^{-2}$ isophote, the so-called isophotal 
radius, is equal to $r_{25} = 354\arcsec = 17.4$~kpc in M63. About 30\% of 
local spiral galaxies have XUV disks as compared to their optical $r_{25}$ 
disks, among which M63 is one of the most extreme cases 
\citep{thilker2005,gildepaz2005,gildepaz2007}. A metallicity gradient of 
$-0.59~{\rm dex}~r_{25}^{-1}$ was reported by \citet{moustakas2010} for the 
calibration of \citet{pilyugin2005}, which gives $12+\log({\rm O/H}) = 
8.59\pm 0.07$ for the central metallicity of M63.

The 21~cm observations of M63 also show the presence of a very large, warped 
gaseous disk extending out to 40~kpc in the major axis radius \citep[e.g.,][]
{battaglia2006}. The warp starts around $r_{25}$ and is exceptionally extended 
and symmetric. The measured H\,I column densities are higher than 
$10^{20}~\rm cm^{-2}$ up to 70\% of the XUV disk. These high H\,I column 
densities plus the extended UV emission, both suggest the presence of 
molecular gas out to large galactocentric radii. Indeed, recent studies have 
demonstrated that strong H\,I emission is on average a good tracer of regions 
rich in molecular gas \citep[e.g.,][]{crosthwaite2002,nieten2006}, and the 
extended UV emission undeniably betrays the presence of relatively young 
stars, hence of the molecular gas necessary for their formation.

Molecular gas was looked for in M63 in the context of the BIMA/SONG survey, the 
first systematic imaging survey of CO(1--0) emission from the centers and disks 
of nearby galaxies \citep[][see their Fig.~43]{helfer2003}. The reported 
high-resolution CO measurements are confined to the very central area of the 
M63 optical disk, $r_{\rm gal} = 96\arcsec = 0.3\,r_{25}$ along the major axis, 
because of the lack of sensitivity of these data at the border of the map with 
a primary beam gain drop by a factor of 2. Single-dish spectra from the FCRAO 
Extragalactic CO Survey \citep[][see their Fig.~87]{young1995}, despite their 
moderate sensitivity and low resolution, show that the CO emission extends over 
a larger area than mapped by BIMA/SONG, reaching a two times larger 
galactocentric radius, but still not extending up to the optical disk limit 
$r_{25}$. The signal measured at $r_{\rm gal} = 180\arcsec$ is equal to 
$\sim 2~\rm K~km~s^{-1}$, which leaves room for a CO line flux decrease 
by a factor of 10 to 20 in the outermost XUV disk (depending on the CO line 
width), when aiming at signals as weak as 10~mK, or weaker, in these extreme 
regions. More recently, \citet{leroy2009} mapped the CO(2--1) line in M63 over 
the full optical disk, as part as the HERACLES survey on the IRAM 30\,m
telescope, and obtained reliable measurements out to $r_{\rm gal} =
0.68\,r_{25}$.

%

\begin{table*}
\caption{CO mapping over the bright UV region at $1.36\,r_{25}$}             
\label{tab:UVregion}      
\centering          
\begin{tabular}{r r r | D{.}{}{2} D{.}{}{3} c | D{.}{}{2} D{.}{}{3} c | c c c}     
\hline\hline 
 & & & \multicolumn{3}{c|}{CO(1--0)} & \multicolumn{3}{c|}{CO(2--1)} & & & \\
\multicolumn{2}{c}{Offsets} & Time & \multicolumn{1}{c}{rms\tablefootmark{a}} & \multicolumn{1}{c}{FWHM\tablefootmark{b}} & $F$\tablefootmark{c} & \multicolumn{1}{c}{rms\tablefootmark{a}} & \multicolumn{1}{c}{FWHM\tablefootmark{b}} & $F$\tablefootmark{c} & $r_{2,1}$\tablefootmark{d} & $M_{\rm H_2}$\tablefootmark{e} & $\Sigma_{\rm H_2}$\tablefootmark{f} \\
($\arcsec$) & ($\arcsec$) & (min) & \multicolumn{1}{c}{(mK)} & \multicolumn{1}{c}{(km~s$^{-1}$)} & (K~km~s$^{-1}$) & \multicolumn{1}{c}{(mK)}                 & \multicolumn{1}{c}{(km~s$^{-1}$)} & (K~km~s$^{-1}$) &                                   & ($\rm 10^6~M_{\sun}$) & ($\rm M_{\sun}~pc^2$) \\
\hline 
$-66$ &   0 &  118 & 5 & 10 & $<0.15$		      & 6 & 10 & $<0.18$ &         & $<0.97$       & $<0.38$ \\
$-66$ & +22 &  117 & 5 & 11 & $0.14\pm 0.06\ddag$ & 6 & 10 & $<0.18$ & $<0.39$ & $0.90\pm0.39$ & $0.35\pm0.15$ \\
$-44$ &   0 &  186 & 4 & 10 & $<0.12$		      & 4 & 10 & $<0.12$ &         & $<0.78$       & $0.30$ \\
$-44$ & +22 &  164 & 4 & 10 & $<0.12$		      & 6 & 10 & $<0.18$ &         & $<0.78$       & $0.30$ \\
$-22$ &   0 &  171 & 4 & 14 & $0.21\pm 0.06 $     & 5 & 10 & $<0.15$ & $<0.22$ & $1.36\pm0.39$ & $0.53\pm0.15$ \\
$-22$ & +22 &  148 & 4 & 10 & $<0.12$		      & 6 & 10 & $<0.18$ &         & $<0.78$       & $<0.30$ \\
   0  &   0 &  164 & 4 & 10 & $<0.12$		      & 5 & 10 & $<0.15$ &         & $<0.78$       & $<0.30$ \\
   0  & +22 &  148 & 5 & 10 & $<0.15$		      & 6 & 10 & $<0.18$ &         & $<0.97$       & $<0.38$ \\
 +22  &   0 &  123 & 5 & 10 & $<0.15$		      & 7 & 10 & $<0.21$ &         & $<0.97$       & $<0.38$ \\
 +22  & +22 &  116 & 5 & 10 & $<0.15$		      & 8 & 10 & $<0.24$ &         & $<0.97$       & $<0.38$ \\
 +44  &   0 &   63 & 7 & 10 & $<0.21$		      & 9 & 10 & $<0.27$ &         & $<1.36$       & $<0.53$ \\
 +44  & +22 &   62 & 7 & 10 & $<0.21$		      & 8 & 10 & $<0.24$ &         & $<1.36$       & $<0.53$ \\
\hline
\multicolumn{2}{c}{all offsets} & 1932 & 1 & 23 & $0.11\pm 0.02$ & 2 & 10 & $<0.06$ & $<0.17$ & $0.71\pm 0.13$ & $0.28\pm 0.05$ \\
\hline
\end{tabular}
\tablefoot{
The last separate line gives the values obtained when summing all scans at the 
12 offset pointings.
\tablefoottext{a}{Rms noises at, respectively, 3~mm and 1~mm, measured per 
channel of $3.2~\rm km~s^{-1}$ and $1.3~\rm km~s^{-1}$.} 
\tablefoottext{b}{Full widths at half maximum of, respectively, CO(1--0) and 
CO(2--1) lines as determined from fitting Gaussian profiles.} 
\tablefoottext{c}{Integrated CO(1--0) and CO(2-1) line fluxes as determined 
from fitting Gaussian profiles. The value marked with $\ddag$ corresponds to a
potentially less reliable measurement, because of the CO line detection at 
barely $3\,\sigma$. Upper limits are $3\,\sigma$ and are calculated assuming a 
$10~\rm km~s^{-1}$ line width.}
\tablefoottext{d}$^+$\tablefoottext{e}$^+$\tablefoottext{f}{CO luminosity 
ratios, H$_2$ masses, and H$_2$ surface densities, respectively, calculated 
using the same prescriptions as in Table~\ref{tab:M63disk}.}
}
\end{table*}
%

\section{Observations and data reduction}
\label{sect:observations}

The CO observations were performed with the IRAM 30\,m millimeter-wave 
telescope at Pico Veleta, Spain, during a first run on September 10--17, 2007 
under poor weather conditions, and during a second run on November 2, 16, and 
27 and December 1, 2007 under good-to-excellent weather conditions. We used 
four single-pixel heterodyne receivers, simultaneously, two centered on the 
$^{12}$CO(1--0) line at 115.271~GHz, and two on the $^{12}$CO(2--1) line at 
230.538~GHz. The telescope half-power beam widths at these two frequencies are 
$22\arcsec$ and $11\arcsec$, respectively. The data were recorded using the 
VESPA autocorrelator with 640~MHz bandwidth and 1.25~MHz resolution at 3~mm, 
and two 1~MHz filter banks (512 channels each) at 1~mm. The resulting 
velocity coverage at 115.271~GHz is $1665~\rm km~s^{-1}$ with a spectral
resolution of $3.2~\rm km~s^{-1}$. The corresponding values at 230.538~GHz are 
$666~\rm km~s^{-1}$ and $1.3~\rm km~s^{-1}$.

First, we performed a CO mapping along the major axis of the M63 disk from 
the center of the galaxy located at (J2000) $\rm RA = 13h\,15m\,49.3s$, 
$\rm Dec = +42^{\circ}\,01m\,45.4s$ out to the galactocentric radius 
$r_{\rm gal} = 572\arcsec$ by aligning 27 single pointings with a spatial 
sampling of $22\arcsec$ in the radial direction sustaining a position angle 
$\rm PA = 105\degr$. Second, we mapped a bright UV region in the outer regions 
of the M63 disk centered on (J2000) $\rm RA = 13h\,15m\,07.0s$, $\rm Dec = 
+42^{\circ}\,00m\,00.0s$ and located at the galactocentric radius 
$r_{\rm gal} = 483\arcsec$ with $6\times 2 = 12$ pointings following the 
sequence from ($-66\arcsec$;~$0\arcsec$) to ($+44\arcsec$;~$+22\arcsec$) 
offsets relative to the central coordinates with a $22\arcsec$ spatial sampling 
in the right-ascension direction and the same in the declination direction. In 
Fig.~\ref{fig:M63image} we show the false-color GALEX image of M63 on top of 
which the regions targeted for CO emission are plotted. Observations were 
performed in wobbler-switching mode with the maximum symmetrical azimuthal 
wobbler throw of $240\arcsec$ allowed, corresponding to 11.8~kpc in projected 
distance for this spiral galaxy located at a distance of 10.1~Mpc. The total 
on-source exposure time obtained per pointing in the M63 disk map and the UV 
region map is listed in Tables~\ref{tab:M63disk} and \ref{tab:UVregion} 
(column~3), respectively. Only scans with a system temperature lower than 400~K 
at 3~mm were retained for analysis.

The data were reduced with the CLASS software from the GILDAS package. All the
spectra obtained with the two receivers tuned on the $^{12}$CO(1--0) line and 
corresponding to scans at the same pointings were summed up without any 
smoothing. On the other hand, the spectra obtained with the two receivers tuned 
on the $^{12}$CO(2--1) line were first Hanning-smoothed to a resolution of 
$2.6~\rm km~s^{-1}$, since the expected average CO line full width half maximum 
is approximately $10~\rm km~s^{-1}$, before being summed up when corresponding 
to scans at the same pointings. No baseline subtraction was performed on 
individual spectra before summing because we simply did a linear sum. In 
Tables~\ref{tab:M63disk} and \ref{tab:UVregion} (columns~4 and 7), we list the 
achieved $1\,\sigma$ rms in mK at 3~mm and 1~mm for each pointing of the M63 
disk map and the UV region map, respectively\footnote{The efficiencies of the 
ABCD receivers on the IRAM 30\,m telescope are $5.9~\rm Jy~K^{-1}$ at 3~mm and 
$7.2~\rm Jy~K^{-1}$ at 1~mm (see the online IRAM wiki pages at 
http://www.iram.es/IRAMES/mainWiki/Iram30mEfficiencies).}. These rms values 
were obtained with a baseline subtraction of degree 0 and with windows set to 
$300-400~\rm km~s^{-1}$, both at 3~mm and 1~mm, as defined by the velocity 
positions of detected CO(1--0) and CO(2--1) lines\footnote{Except for the very
few inner pointings at $r_{\rm gal} = 0$, $22\arcsec$, and $44\arcsec$, for 
which a larger window was set, given the larger widths of the CO(1--0) and 
CO(2--1) lines.}. At 3~mm we obtained on average rms values between 4~mK and 
21~mK in the M63 disk map, while for seven pointings around the isophotal 
radius, $r_{25}$, and for all the pointings of the UV region map, we pushed 
the rms limit down to 4~mK to 7~mK.

%

\section{Results}
\label{sect:results}

As summarized in Sect.~\ref{sect:M63}, the CO(1--0) and CO(2--1) lines were 
previously mapped over the full optical disk of M63 in the context of various 
CO surveys of nearby galaxies. However, none of these observations reaches the 
sensitivity of our data, in particular around the isophotal radius, $r_{25}$, 
and beyond. High sensitivity can be achieved by mapping CO through individual 
beam pointings, which is a technique that proves to be very efficient for 
obtaining deep observations of specific, not extended, areas of a galaxy. 

%

\begin{figure*}[!]
\centering
\includegraphics[width=\textwidth]{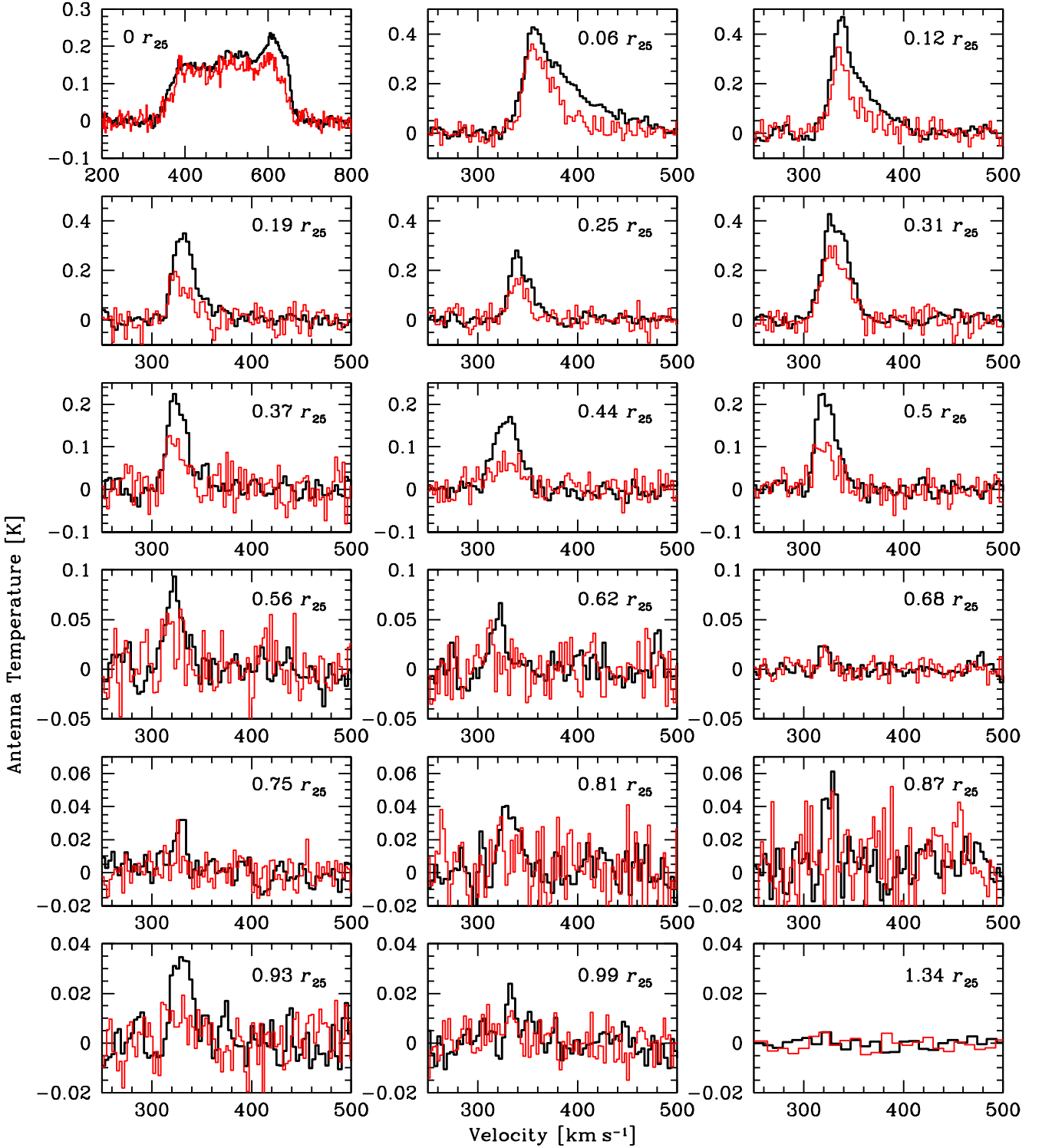}
\caption{Mosaic of CO spectra obtained when mapping the M63 disk along its 
major axis from the center of the galaxy out to the isophotal radius, $r_{25}$.
The CO(2--1) spectra (red thin line) are overplotted on the CO(1--0) spectra
(black thick line). While CO(1--0) is clearly detected up to the optical 
radius, CO(2--1) appears to be confined to $r_{\rm gal} \lesssim 0.68\,r_{25}$. 
The bottom right hand panel shows the CO(1--0) and CO(2--1) spectra obtained 
when summing all scans at the ten outermost pointings from $r_{\rm gal} = 
1.06\,r_{25}$ to $1.6\,r_{25}$ and smoothed to a resolution of 
$9.7~\rm km~s^{-1}$ and $10.4~\rm km~s^{-1}$, respectively. No CO emission is 
detected at these outermost pointings.}
\label{fig:stampM63disk}
\end{figure*}
%

\begin{figure}
\centering
\includegraphics[width=9cm]{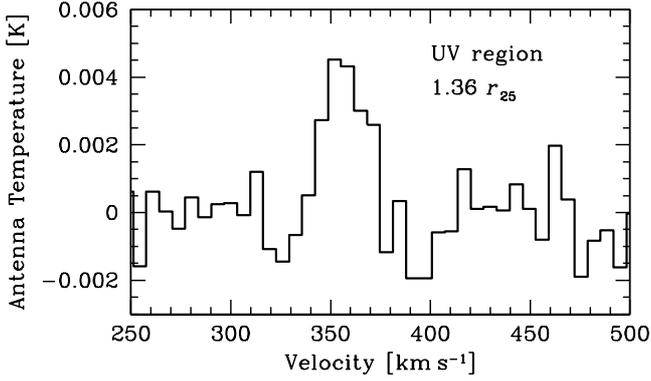}
\caption{CO(1--0) spectrum obtained by summing up all spectra at the 12 
pointings used to map the bright UV region at $r_{\rm gal} = 1.36\,r_{25}$. 
There is a clear CO emission detection in this region of M63 well beyond the 
optical radius. CO(2--1) remains undetected.}
\label{fig:CO-UVregion}
\end{figure}

The results of our CO emission mapping of M63 along the major axis of its disk 
and over the bright UV region at $r_{\rm gal} = 1.36\,r_{25}$ are presented in 
Tables~\ref{tab:M63disk} and \ref{tab:UVregion}, respectively. We provide the 
CO(1--0) and CO(2--1) line full widths half maximum, FWHM (columns~5 and 8) and 
the integrated CO(1--0) and CO(2--1) line fluxes (columns~6 and 9) at each 
pointing of the mapping, as determined from fitting Gaussian functions to the 
CO profiles obtained by summing up all spectra corresponding to all scans per 
pointing. The CO(1--0) and CO(2--1) line FWHM are in the range from 
$9~\rm km~s^{-1}$ to $53~\rm km~s^{-1}$ and from $6~\rm km~s^{-1}$ to 
$39~\rm km~s^{-1}$, respectively. They are greater than those expected for 
individual giant molecular clouds (GMCs) that have typical line widths of 
$10.4~\rm km~s^{-1}$ \citep{solomon1987}, except for a few pointings around 
$r_{25}$ and over the UV region. This implies that mostly an ensemble of 
molecular clouds is emitting per beam of 1~kpc and 0.5~kpc at 3~mm and 1~mm, 
respectively. In Tables~\ref{tab:M63disk} and \ref{tab:UVregion} we also list 
the molecular gas masses per pointing (column~11), as derived using the 
``standard'' Galactic CO(1--0)-to-H$_2$ conversion factor $X_{\rm CO} = 
2\times 10^{20}~\rm cm^{-2}(K~km~s^{-1})^{-1}$ \citep{dickman1986} and 
including a correction factor of 1.36 for helium. The inferred masses are in 
the range $M_{\rm H_2} = (0.9-322)\times 10^6~\rm M_{\sun}$. 
Figure~\ref{fig:stampM63disk} shows the summed-up spectra of all scans 
{\it per} pointing of the M63 disk map along the major axis, while 
Fig.~\ref{fig:CO-UVregion} shows the resulting spectrum obtained by summing up 
all spectra of all scans at the 12 pointings used to map the external UV 
region. 

%

%

The radial mapping (Fig.~\ref{fig:stampM63disk}) clearly shows a detection of 
the CO(1--0) emission out to the galactocentric radius $r_{\rm gal} = 
352\arcsec$, namely out to the isophotal radius $r_{25} = 354\arcsec = 
17.4~\rm kpc$. 
A severe drop in the CO flux is nevertheless observed as a function of the 
galactocentric radius. Beyond $r_{25}$ and out to $1.6\,r_{25}$, the limit of 
our CO search along the M63 major axis, no CO emission is detected anymore 
(Table~\ref{tab:M63disk}). Even by summing up all the spectra corresponding to 
all scans at the ten outermost pointings and by smoothing the resulting 
spectrum to a resolution of $9.7~\rm km~s^{-1}$, i.e., similar to the typical 
FWHM of CO(1--0) lines detected at the outermost pointings, no CO emission 
is observed. We do, however, derive a stringent $3\,\sigma$ upper limit on the 
integrated CO(1--0) line flux at $r_{\rm gal} > r_{25}$ of 
$F_{\rm CO(1-0)} < 0.06~\rm K~km~s^{-1}$, when assuming a line width of 
$10~\rm km~s^{-1}$. This is equivalent to a molecular gas mass of $M_{\rm H_2} 
< 3.9\times 10^5~\rm M_{\sun}$. On the other hand, the CO(2--1) emission is 
securely detected only out to $r_{\rm gal} = 242\arcsec$, i.e., $0.68\,r_{25}$, 
in agreement with the CO(2--1) mapping by \citet{leroy2009}, while there are 
two very tentative detections at $2\,\sigma$ at the pointings $r_{\rm gal} = 
0.93\,r_{25}$ and $r_{25}$ (Table~\ref{tab:M63disk}). The CO(2--1) line thus 
appears to be excited over about 2/3 of the optical disk. The corresponding CO 
luminosity ratios, $r_{2,1} = L'_{\rm CO(2-1)}/L'_{\rm CO(1-0)}$, 
can be found in Table~\ref{tab:M63disk} (column~10). They vary randomly between 
0.13 and 0.24 (in the center of M63) along the galactocentric radius, without 
showing any radial evolution. The absence of any clear detection of the 
CO(2--1) line beyond $r_{\rm gal} = 0.68\,r_{25}$ suggests that the subthermal 
excitation sets in around this galactocentric radius. However, this has to be 
considered with caution because the major axis pointings probe only a small 
portion of the M63 disk at each radius and the \citet{leroy2009} CO(2--1) map 
of the entire M63 disk is not very deep and because there seems still to be 
plenty of star-forming regions at $r_{\rm gal} > 0.68\,r_{25}$ with high enough 
densities (locally) to thermalize the CO(2--1) line (see 
Fig.~\ref{fig:M63image}). 

Whereas our survey for CO emission along the major axis of the M63 disk seems 
to indicate that the molecular gas is confined to the isophotal radius $r_{25}$ 
of the galaxy, we find a convincing CO(1--0) detection in the selected bright 
UV region at $r_{\rm gal} = 1.36\,r_{25}$. The sum of all spectra corresponding
to all scans at the 12 pointings used to map the UV region leads to a CO(1--0) 
detection at $5.5\,\sigma$ with an integrated line flux $F_{\rm CO(1-0)} = 
0.11\pm 0.02~\rm K~km~s^{-1}$ (Fig.~\ref{fig:CO-UVregion}). This corresponds to 
a molecular gas mass of $M_{\rm H_2} = 7.1\times 10^5~\rm M_{\sun}$. The 
CO(1--0) emission appears to be maximal at the offset 
($-22\arcsec$;~$0\arcsec$), where it is detected at $3-4\,\sigma$ 
(Table~\ref{tab:UVregion}). The CO(2--1) emission is, on the other hand, not 
detected down to an integrated line flux $F_{\rm CO(2-1)} < 
0.06~\rm K~km~s^{-1}$, obtained when summing up all the spectra corresponding 
to all scans at the 12 pointings.

%

\section{Discussion}
\label{sect:discussion}

A few hundred nearby galaxies have been the object of intensive surveys for
molecular gas \citep[e.g.,][]{young1991,young1995,helfer2003,leroy2009}. 
However, for only a very few have deep CO searches in their outer disk regions 
been undertaken. Here, we show the existence of molecular gas up to the 
isophotal radius, $r_{25}$, and beyond in the nearby spiral galaxy M63. This is 
the fourth such a detection of molecular gas in the outskirts after the 
spiral galaxies NGC\,4414 \citep{braine2004}, NGC\,6946 \citep{braine2007}, and 
M33 \citep{braine2010}. To analyze the impact of our observations in the 
context of star formation in the outer disk of M63, we present the radial 
profiles of our CO observations along with complementary data of other star 
formation tracers (FUV, NUV, H$\alpha$, 24~$\mu$m) and H\,I, and we investigate 
the behavior of the Kennicutt-Schmidt relations across the galaxy and beyond 
$r_{25}$ in the bright UV region. Characterizing the Kennicutt-Schmidt relation 
in the outer disk regions, namely in environments with low metallicities, gas 
temperatures, excitations, and gas densities, which are all properties hostile 
to star formation, is our main interest.

%

\begin{table}
\caption{Public ancillary data of M63}            
\label{tab:literature-data}      
\centering          
\begin{tabular}{llcl}     
\hline\hline 
Band or line & Telescope & FWHM        & Reference \\
             &           & ($\arcsec$) & \\
\hline 
CO(2--1)     & IRAM    & 13.4 & \citet{leroy2009} \\
FUV          & GALEX   & 4.3  & \citet{gildepaz2007} \\
NUV          & GALEX   & 5.3  & \citet{gildepaz2007} \\
H$\alpha$    & KPNO    & 0.38 & \citet{kennicutt2008} \\
24~$\mu$m    & Spitzer & 6    & \citet{dale2009} \\
H\,I (21~cm) & VLA     & 6    & \citet{walter2008} \\
\hline
\end{tabular}
\end{table}
%

\begin{figure*}
\centering
\includegraphics[width=1.0\textwidth]{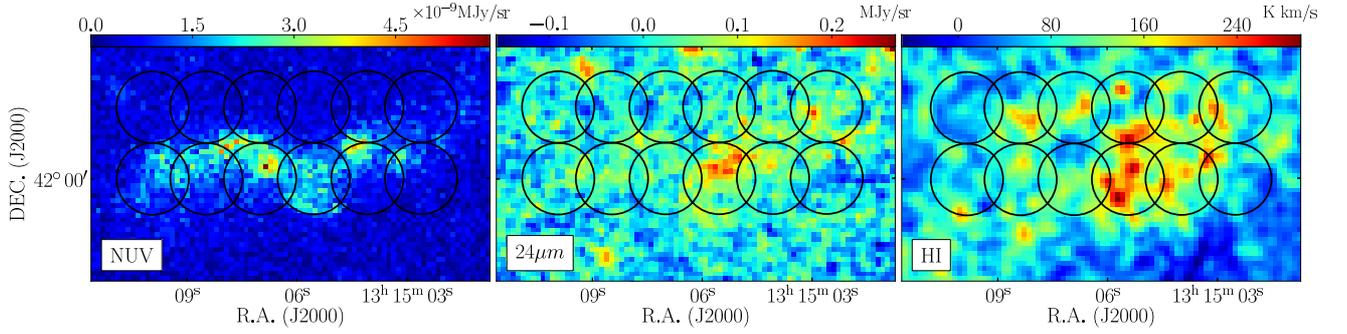}
\caption{Multiwavelength views of the bright UV region at $r_{\rm gal} = 
1.36\,r_{25}$, showing from left to right the NUV, 24~$\mu$m, and H\,I 21~cm 
emission in flux units of $10^{-9}~\rm MJy~sr^{-1}$, $\rm MJy~sr^{-1}$, and 
$\rm K~km~s^{-1}$, respectively, as indicated by the color bars. The circles of 
$22\arcsec$ diameter refer to the positions of the 12 pointings used to map the 
CO(1--0) emission. The H$\alpha$ emission is undetected in this external region 
of the M63 galaxy.}
\label{fig:UVregion-multipanel}
\end{figure*}
%

\begin{figure*}
\centering
\includegraphics[width=1.0\textwidth]{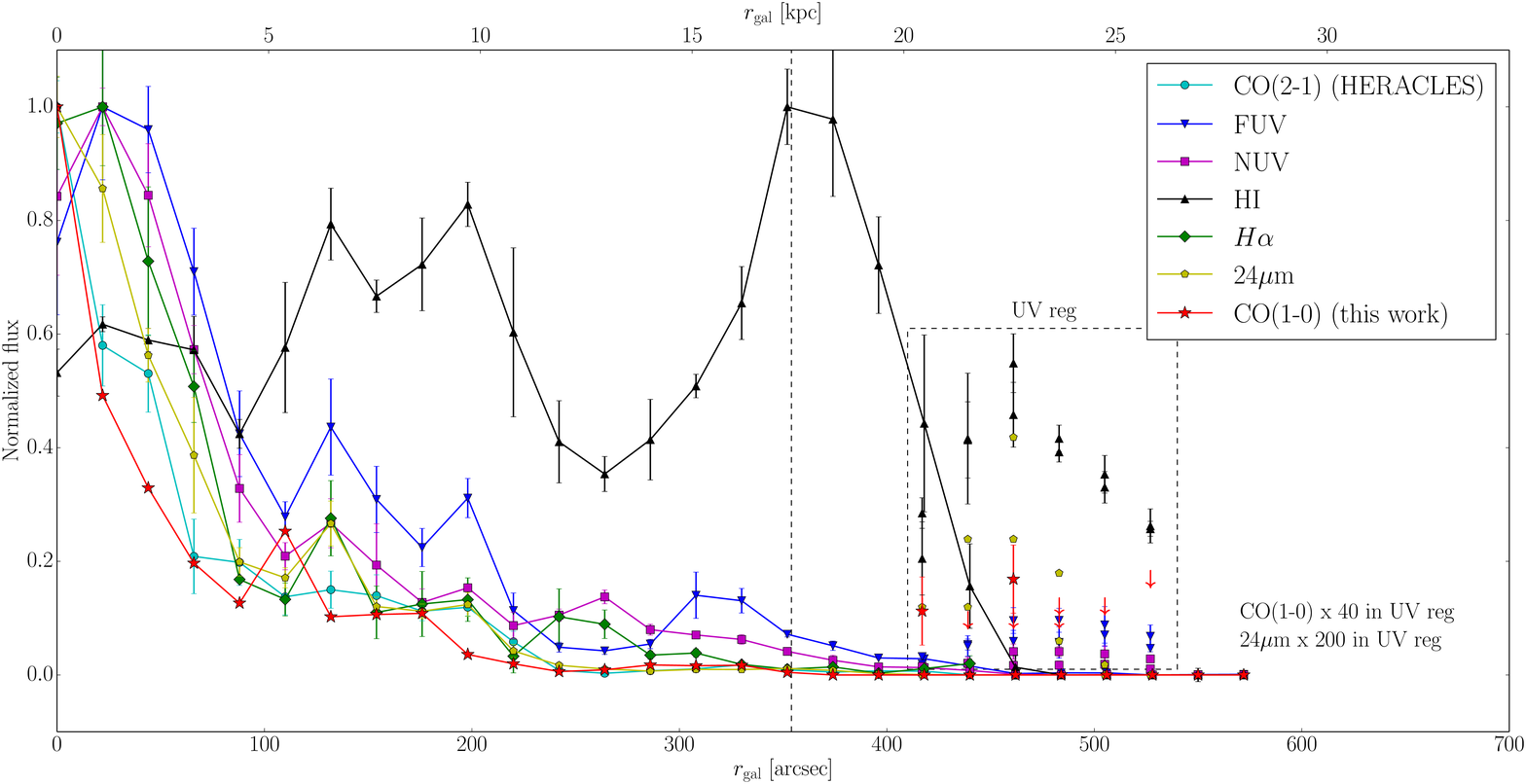}
\caption{Radial profiles of FUV, NUV, H$\alpha$, 24~$\mu$m, different star 
formation tracers, and H\,I taken from the literature (see 
Table~\ref{tab:literature-data}), as measured along the major axis of the M63 
disk, which we compare to our CO(1--0) radial profile (red stars) and to 
the CO(2--1) radial profile from the HERACLES survey \citep{leroy2009}. The 
measured fluxes are plotted in a normalized flux scale in order to allow 
relative comparisons. All the star formation tracers nicely follow the spiral 
arm structure at $r_{\rm gal} \simeq 130$, 200, and probably $260\arcsec$. We 
observe a severe drop with the galactocentric radius for all star formation 
tracers, as well as for CO, such that close to the $r_{25}$ limit (dashed line) 
and beyond all tracers and CO are practically absent with the exception of UV 
and H\,I. The CO emission is again detected in the external bright UV region. 
The measurements obtained in this UV region are shown in the dashed box at the 
correct galactocentric radii. The CO(1--0) and 24~$\mu$m fluxes are multiplied 
by a factor of 40 and 200, respectively, to make them visible. The double data 
points plotted at each $r_{\rm gal}$ correspond to the parallel pointings used 
to map the UV region (see Table~\ref{tab:UVregion}). This external UV region 
has relatively high FUV and NUV emissions and is dominated by H\,I.}
\label{fig:radial-profiles}
\end{figure*}
%

\subsection{Radial profiles}
\label{sect:radialprofiles}

To compare the CO distribution in M63 along the major axis of its disk to H\,I 
and other star formation tracers' distributions, we built comparative radial 
profiles of the CO, H\,I, FUV, NUV, H$\alpha$, and 24~$\mu$m emission. 
Altogether they provide a complete view of the past and ongoing star formation 
that took or takes place in the galaxy and of the ingredients needed to sustain 
the star formation: molecular and atomic gas. The respective data are taken 
from the literature and are described in Table~\ref{tab:literature-data}. In 
Fig.~\ref{fig:UVregion-multipanel} we show the multiwavelength views of the 
bright UV region at $r_{\rm gal} = 1.36\,r_{25}$ with, from left to right, the 
NVU, 24~$\mu$m, and H\,I 21~cm emission.

As these data were obtained with different instruments and hence at various 
FWHM resolutions, we first convolved all these data to the resolution of our 
CO(1--0) data which is equal to the half-power beam width of the IRAM 30\,m 
telescope at 3~mm, i.e., $22\arcsec$. The images were then rotated by $20\degr$ 
in a north-to-west direction to have the major axis in the horizontal axis, and 
deprojected by correcting for a $55\degr$ inclination angle to bring the galaxy 
face-on. Finally, we did aperture photometry on these processed images with 
apertures of $22\arcsec$ diameter using the `qphot' task of the 
`digiphot.apphot' package of IRAF\footnote{IRAF is distributed by the National 
Optical Astronomy Observatories, which are operated by the Association of 
Universities for Research in Astronomy, Inc., under cooperative agreement with 
the National Science Foundation.}, and we subtracted an average sky value 
measured away from the galaxy. The photometric measurements were performed at 
the 27 positions of the radial cut along the M63 major axis used for the CO 
radial mapping, as well as at the 12 positions used to map CO in the external 
bright UV region, following the same spacing and pointings exactly.

The resulting CO(1--0), CO(2--1), FUV, NUV, H\,I, H$\alpha$, and 24~$\mu$m 
radial profiles along the M63 radial cut are shown in 
Fig.~\ref{fig:radial-profiles} (connected data points), where the measured 
fluxes are plotted in a normalized flux scale to allow relative comparisons. 
All the profiles nicely follow the spiral structure observed at galactocentric 
radii $r_{\rm gal} \simeq 130$, 200, and probably $260\arcsec$. They all show a 
severe drop with the galactocentric radius, except the profile of the atomic 
gas, which shows a completely different behavior. We observe that close to the 
$r_{25}$ limit, the star formation tracers and CO begin to vanish considerably, 
and beyond the $r_{25}$ limit, they all are practically absent with the 
exception of the faint UV emission. The H\,I emission is very strong beyond 
the $r_{25}$ limit. This shows the importance of looking for molecular gas 
beyond the optical limit, where evidence of star formation is clearly brought by the UV emission. 

%

\begin{figure}
\centering
\includegraphics[width=9.2cm]{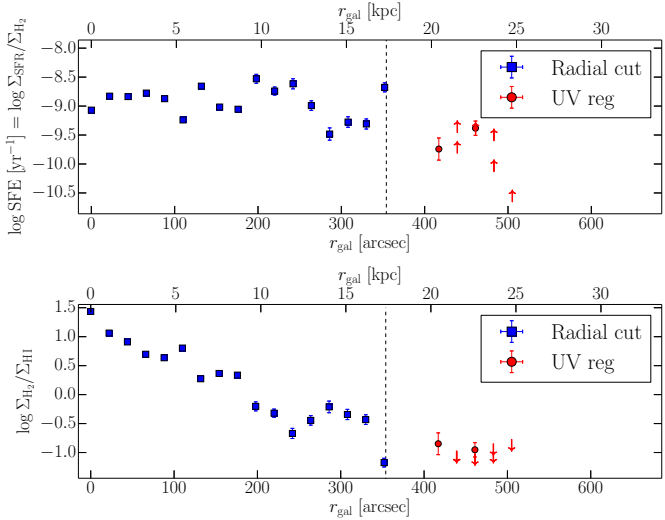}
\caption{Radial profiles of the star formation efficiency (upper panel) and the 
H$_2$-to-H\,I surface density ratio (bottom panel). The SFE is defined here as 
the star formation rate per unit of molecular gas. The blue squares correspond 
to the 17 pointings used to map CO along the M63 major axis (the radial cut) 
out to the isophotal radius, $r_{25}$ (dashed line), and the red circles 
correspond to the 8 pointings with 24~$\mu$m emission detection used to map CO 
over the bright UV region at $r_{\rm gal} = 1.36\,r_{25}$. The double data 
points plotted for the UV region at each $r_{\rm gal}$ correspond to the 
parallel pointings used to map this region (see Table~\ref{tab:UVregion}).}
\label{fig:SFE-H2HI}
\end{figure}
%

While the CO emission is not detected beyond the $r_{25}$ limit in the M63 
major axis map, it is again detected in the external UV region at 
$r_{\rm gal} = 1.36\,r_{25}$. The corresponding CO(1--0) fluxes, multiplied by 
a factor of 40 to make them visible (showing the depth of our data), with 
H\,I and other star formation tracer fluxes as measured in the UV region, are 
also shown in Fig.~\ref{fig:radial-profiles}. The FUV and NUV emissions are 
observed in all pointings over the external UV region, while the H$\alpha$ 
emission is undetected and a faint 24~$\mu$m emission, whose fluxes were 
multiplied by a factor of 200 to make them visible, is observed in only eight 
pointings (see also Fig.~\ref{fig:UVregion-multipanel}). It is interesting to 
point out that the FUV and NUV fluxes in the external UV region are similar  
to those at the pointing at $r_{25}$, the galactocentric limit where CO is 
still detected along the M63 major axis. Beyond $r_{25}$, the FUV and NUV 
fluxes decrease and get weaker than in the external UV region. This suggests a 
tight correlation between the UV flux, namely the intensity of star formation 
and the CO flux that should trace the amount of molecular gas, since the CO 
emission starts to be detected at a given UV flux level. If correct, the CO 
emission is not detected beyond $r_{25}$ along the major axis mapping simply 
because the amount of CO, corresponding to the molecular gas needed to sustain 
the star formation that is taking place there, falls below our detection 
threshold. Consequently, it is likely that molecular gas is present in the 
outer regions of the M63 disk at least as long as the XUV is present.

%

\begin{table*}
\caption{Best-fitting bisector linear Kennicutt-Schmidt relations obtained 
for M63 in log space}         
\label{tab:fitting-results}      
\centering    
\renewcommand{\arraystretch}{1.2}    
\begin{tabular}{cccccccccccc}
\hline
\hline 
& \multicolumn{7}{ c }{{This work}}& &\multicolumn{3}{ c }{{Bigiel et~al.\ 2008}} \\ \cline{2-8} \cline{9-12}
& \multicolumn{3}{ c }{Radial cut out to $r_{25}$} & &  \multicolumn{3}{ c }{UV region at $1.36\,r_{25}$} & & \multicolumn{3}{ c }{Radial cut out to $0.68\,r_{25}$} \\ \cline{2-4} \cline{6-8} \cline{10-12}
& \multicolumn{1}{ c }{$\log A$} & \multicolumn{1}{ c }{$N$} & \multicolumn{1}{ c }{$\sigma$}& & \multicolumn{1}{ c }{$\log A$} & \multicolumn{1}{ c }{$N$}& \multicolumn{1}{ c }{$\sigma$}& &\multicolumn{1}{ c }{$\log A$} & \multicolumn{1}{ c }{$N$}& \multicolumn{1}{ c }{$\sigma$}\\ \cline{1-12}
\multicolumn{1}{ c }{$\rm H_2$}  &
\multicolumn{1}{ c } {-1.93} & {1.06} & {0.29} & & \multicolumn{1}{ c }{0.17$\dag$} & {3.0$\dag$} & {--} & & \multicolumn{1}{ c }{-2.22} & \multicolumn{1}{ c }{0.92}  & \multicolumn{1}{ c }{0.10}    \\ 
\multicolumn{1}{ c }{$\rm H\,I+H_2$}  &
\multicolumn{1}{ c } {-2.43} & {1.73} & {0.31} & & \multicolumn{1}{ c }{-2.20} & {4.7} & {0.35} & & \multicolumn{1}{ c }{-2.63} & \multicolumn{1}{ c }{1.58}  & \multicolumn{1}{ c }{0.22}   \\ \cline{1-12}
\end{tabular}
\tablefoot{The Kennicutt-Schmidt relations are best parametrized by power laws 
of the form $\Sigma_{\rm SFR} = A(\Sigma_{\rm gas})^N$, which in log space 
translates into linear relations. $\dag$ Only indicative $\log A$ and $N$ 
values, since they rely only on two data points, which makes them very 
uncertain.}
\end{table*}
%

The considerably larger amount of H\,I observed in the external UV region 
relative to CO leads us to speculate that H\,I is more likely a precursor of 
H$_2$ rather than a product of photodissociation due to UV radiation. However, 
the situation might be more complicated in reality. The main complication we 
may invoke is the reliability of CO as a proxy of H$_2$. As discussed in 
Sect.~\ref{sect:results} and shown in Fig.~\ref{fig:radial-profiles}, the 
absence of a clear CO(2--1) line detection beyond $r_{\rm gal} = 0.68\,r_{25}$, 
with a CO luminosity ratio, $r_{2,1}$, below $\sim 0.1$ (see 
Table~\ref{tab:M63disk}), may betray a decrease in the excitation temperature 
with the galactocentric radius. If this is the case, it may be that even if 
substantial H$_2$ is present in the outer regions of the M63 disk, the CO lines 
may be quite weak, hence remain undetected.

In Fig.~\ref{fig:SFE-H2HI} we plot the radial profiles of the star formation 
efficiency (SFE) and the H$_2$-to-H\,I surface density ratio. The SFE is the 
star formation rate (SFR) per unit of molecular gas or the inverse of the gas 
depletion timescale, i.e., the time required for present-day star formation to 
consume the gas reservoir. It is calculated here as ${\rm SFE} = 
\Sigma_{\rm SFR}/\Sigma_{\rm H_2}$. The various surface densities are computed 
using Eqs.\ (\ref{sigmaSFR}) to (\ref{sigmaH2}). 
In the radial cut we only consider the pointings up to the isophotal radius, 
$r_{25}$, where we obtained convincing CO detections (17 pointings), and in the 
UV region we retain pointings with 24~$\mu$m detections only (8 pointings) in 
order to accurately estimate the star formation rate density. We 
observe that the SFE, as measured over the radial cut of M63, is roughly 
constant as a function of the galactocentric radius out to the $r_{25}$ limit. 
This is in line with the findings by \citet{leroy2008}, which we now extend out 
to the isophotal radius. Beyond $r_{25}$, the two SFE detections tend to show a 
drop in the SFE in the bright UV region, but upper limits in the other data 
points prevent us from drawing definitive conclusions. The H$_2$-to-H\,I 
surface density ratio shows a smooth decrease with the galactocentric radius, 
such that the inner regions of the M63 disk are H$_2$-dominated, while the 
regions at $r_{\rm gal} > 0.5\,r_{25}$ end up to be H\,I-dominated. This 
transition between a ``mostly-H$_2$'' and a ``mostly-H\,I'' interstellar medium 
(ISM) is found to be a well-defined function of local conditions according to 
\citet{leroy2008}, occurring at characteristic galactocentric radius, stellar 
and gas surface densities, hydrostatic gas pressure, and orbital timescale. At 
the $r_{25}$ limit and beyond, we observe a further and more severe drop of the 
H$_2$-to-H\,I surface density ratio. Nevertheless, the behavior of the radial 
profiles of the SFE and the H$_2$-to-H\,I surface density ratio may well be 
inverted, if the trend toward an increasing CO-to-H$_2$ conversion factor with 
the galactocentric radius, as shown by \citet[][see their Fig.~7]
{sandstrom2013} for M63 over the radial cut out to $r_{\rm gal} \sim 0.7\,r_{25}$, is confirmed.

%

\subsection{Kennicutt-Schmidt relations}

Our acquired CO measurements offer the opportunity to analyze the 
Kennicutt-Schmidt relations across the optical disk of the M63 galaxy and, in 
particular, close to and beyond the isophotal radius, $r_{25}$. This exercise 
is especially of interest in the bright UV region located in the outer disk of 
M63 at the galactocentric radius $r_{\rm gal} = 1.36\,r_{25}$, whose stellar 
emission is solely dominated by the UV emission, so where the star formation 
rate, metallicity, gas temperature, excitation, and gas density are potentially 
lower. For this purpose, we use the photometric measurements obtained in 
Sect.~\ref{sect:radialprofiles} for the FUV, 24~$\mu$m, and H\,I data. At each 
pointing of the CO radial mapping and the CO mapping of the bright UV 
region, we compute the star formation rate surface density, $\Sigma_{\rm SFR}$, 
the atomic gas surface density, $\Sigma_{\rm H\,I}$, and the molecular gas 
surface density, $\Sigma_{\rm H_2}$. We can question how these surface 
densities determined over single pointings along the radial major axis cut are 
representative of surface densities that \citet{leroy2008} obtained by 
performing azimuthal averages over the entire disk of M63. The comparison of 
the respective molecular gas surface densities as a function of the 
galactocentric radius is excellent. The SFR surface densities for the radial 
cut show an offset toward higher values, but the respective $\Sigma_{\rm SFR}$ 
radial profiles are the same. Only the atomic gas surface density measurements 
seem to severely diverge between the radial cut and the azimuthal averages. 
This difference comes from the fact that the radial cut along the major axis 
crosses a spiral arm like structure, which is very bright in H\,I and covers an 
H\,I emission peak in the outer parts of the radial profile. This certainly 
also explains the higher $\Sigma_{\rm SFR}$ we observe for the radial cut, 
where the star formation activity is slightly enhanced.

The SFR surface density is determined using the calibration from
\citet{leroy2008}:
\begin{eqnarray}\label{sigmaSFR}
\Sigma_{\rm SFR}({\rm M_{\sun}~yr^{-1}~kpc^{-2}}) & = & \nonumber \\
8.1\times 10^{-2} F_{\rm FUV}({\rm MJy~sr^{-1}}) + 
3.2\times 10^{-3} F_{24\mu\rm m}({\rm MJy~sr^{-1}}), & &
\end{eqnarray}
which includes the FUV flux to measure the unobscured star formation through 
the emission of O and B stars, and the 24~$\mu$m flux that traces the 
obscured FUV emission that is re-emitted in the far-IR by dust grains. 
The atomic gas surface density is measured with the calibration of 
\citet{bigiel2010}:
\begin{equation}\label{sigmaHI}
\Sigma_{\rm H\,I}({\rm M_{\sun}~pc^{-2}}) = 
0.020 F_{\rm H\,I}({\rm K~km~s^{-1}}),
\end{equation}
which includes a factor of 1.36 to reflect the presence of helium. Finally, 
the molecular gas surface density (see Tables~\ref{tab:M63disk} and 
\ref{tab:UVregion}) is calculated with our integrated CO(1--0) line fluxes 
using the calibration of \citet{leroy2008}:
\begin{equation}\label{sigmaH2}
\Sigma_{\rm H_2}({\rm M_{\sun}~pc^{-2}}) = 
4.4 \cos(i) F_{\rm CO(1-0)}({\rm K~km~s^{-1}}),
\end{equation}
where the ``standard'' Galactic CO-to-H$_2$ conversion factor, $X_{\rm CO} = 
2\times 10^{20}~\rm cm^{-2}~(K~km~s^{-1})^{-1}$ \citep{dickman1986}, as well 
as the factor of 1.36 to account for helium are adopted. The inclination 
correction by an angle $i = 55\degr$ is taken into account. This correction 
is also considered in the SFR and atomic gas surface densities when 
deprojecting the FUV, 24~$\mu$m, and H\,I images (see 
Sect.~\ref{sect:radialprofiles}).

\begin{figure*}[!]
\centering
\includegraphics[width=1.0\textwidth]{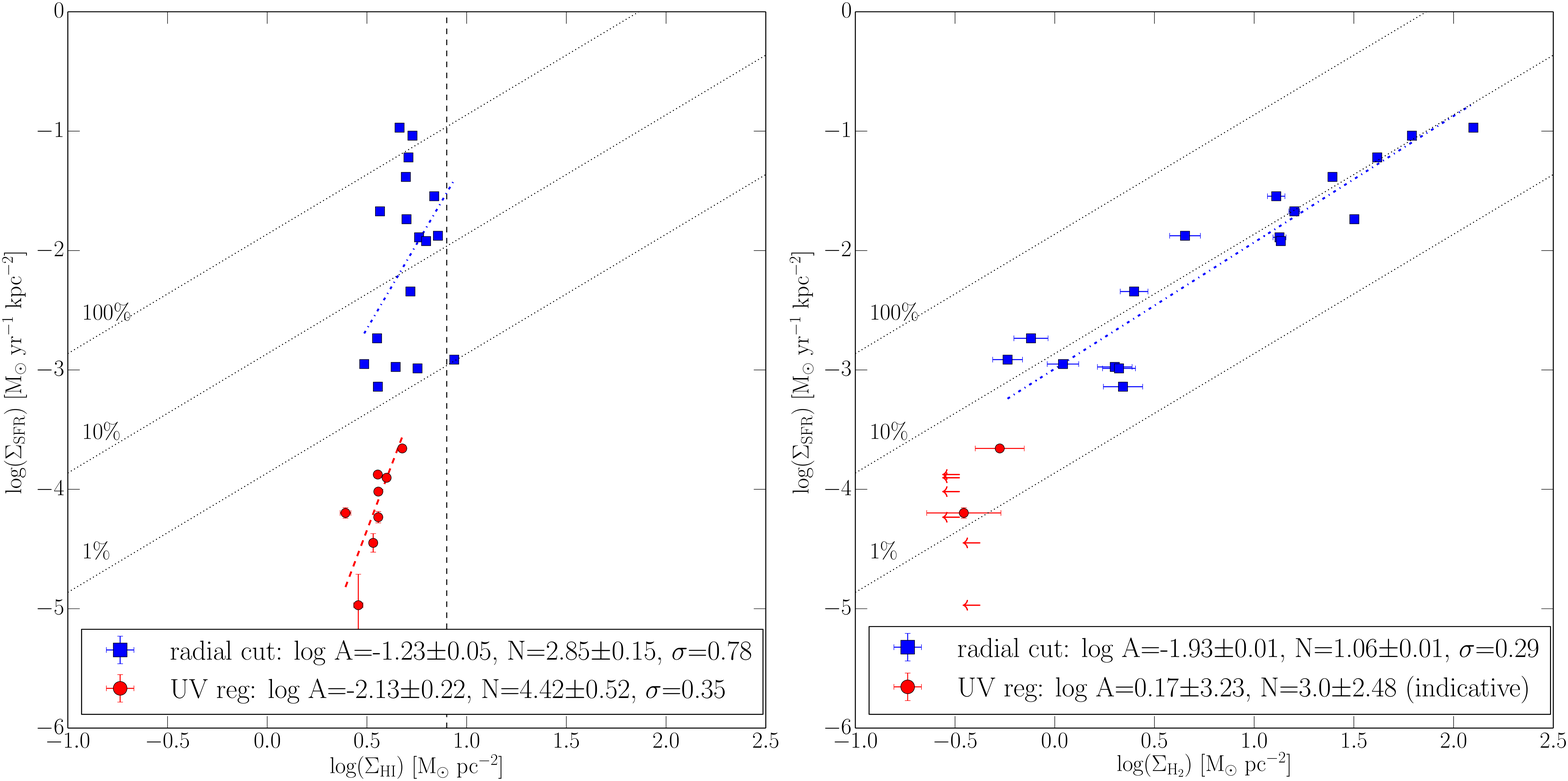}
\includegraphics[width=0.7\textwidth]{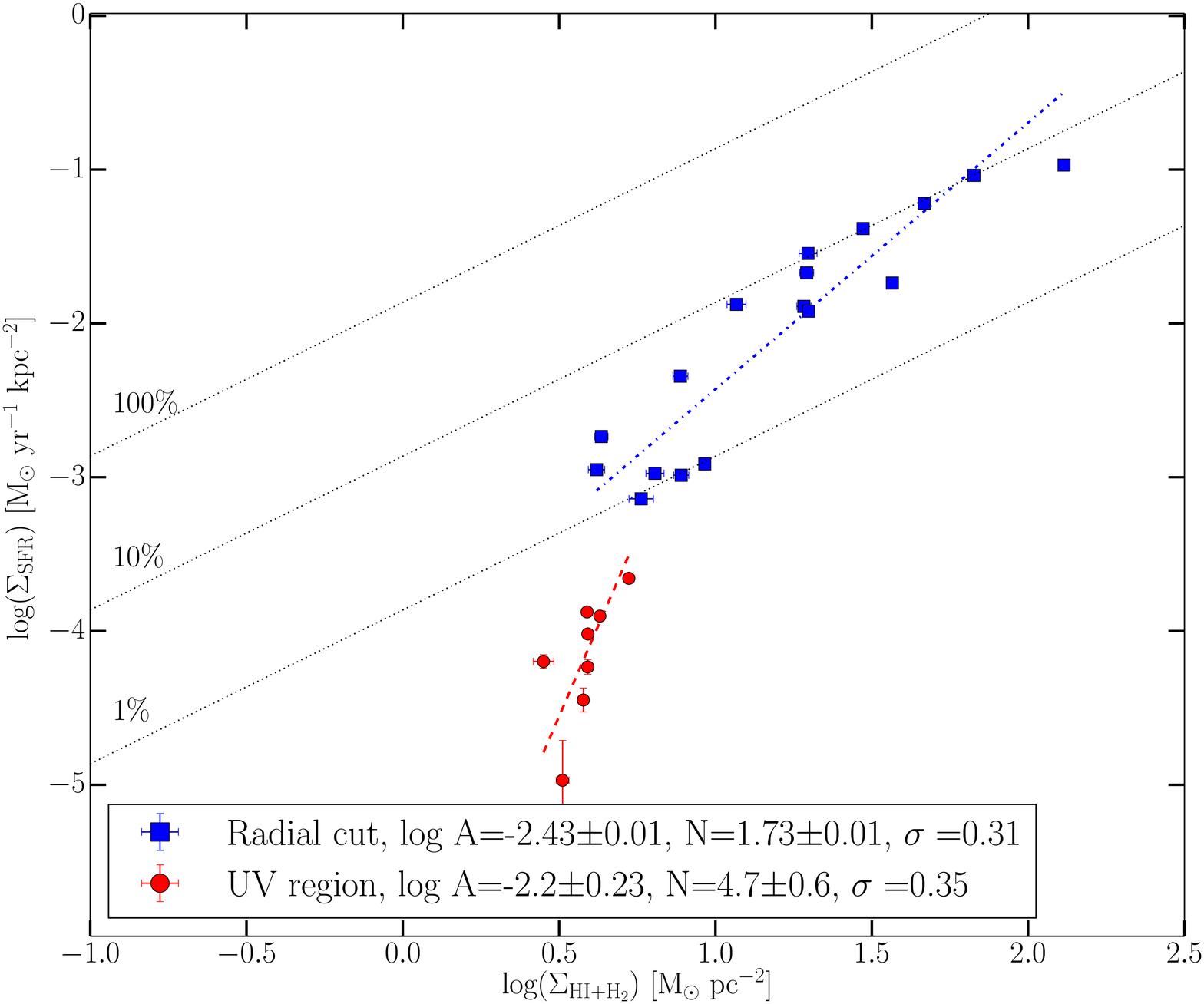}
\caption{Kennicutt-Schmidt relations that relate the SFR surface density to 
the gas surface density plotted for $\Sigma_{\rm H\,I}$ (top-left panel), 
$\Sigma_{\rm H_2}$ (top-right panel), and $\Sigma_{\rm H\,I+H_2}$ (bottom 
panel). The blue squares correspond to the 17 pointings used to map CO along 
the M63 major axis (the radial cut) out to the isophotal radius, $r_{25}$, and 
the red circles correspond to the 8 pointings with 24~$\mu$m emission detection 
used to map CO over the bright UV region at $r_{\rm gal} = 1.36\,r_{25}$. The 
K-S relation is best parametrized by a power law of the form $\Sigma_{\rm SFR} 
= A (\Sigma_{\rm gas})^N$, which in log space translates into a linear
relation. The best-fitting bisector linear K-S relations are derived separately 
for the radial cut (blue dashed-dotted lines) and for the UV region (red dashed 
lines). The corresponding coefficients $\log A$ and slopes $N$ can be found in 
the labels of each panel. The black dotted lines represent ``isochrones'' of 
constant star formation efficiencies, indicating the level of 
$\Sigma_{\rm SFR}$ needed to consume 100\%, 10\%, and 1\% of the total amount 
of gas within $10^8$ years. The vertical dashed line in the top-left panel 
corresponds to the $\Sigma_{\rm H\,I}\sim 9~\rm M_{\sun}~pc^{-2}$ threshold at 
which the atomic gas saturates \citep{bigiel2008}.}
\label{fig:K-S_laws}
\end{figure*}

We compare the star formation rate surface densities with the atomic and 
molecular gas surface densities separately in the two top panels of 
Fig.~\ref{fig:K-S_laws}. 
The two plots show that the gas surface density of M63 in its radial cut is 
dominated by the molecular gas, whereas the bright UV region is strongly 
dominated by the atomic gas, as also observed in Fig.~\ref{fig:SFE-H2HI} 
(bottom panel). Interestingly, the H\,I surface density of the UV region is 
very similar to the one observed along the radial cut, in contrast to its 
molecular gas surface density, which is significantly lower. Finally, in the 
bottom panel of Fig.~\ref{fig:K-S_laws} we show the star formation rate surface 
density as a function of the total gas surface density, 
$\Sigma_{\rm H\,I+H_2}$, namely the sum of both $\Sigma_{\rm H\,I}$ and 
$\Sigma_{\rm H_2}$.

To characterize the star formation in M63, we use the Kennicutt-Schmidt (K-S) 
relation that relates the SFR surface density to the gas surface density. In 
all plots of Fig.~\ref{fig:K-S_laws}, we fit a power law of the form 
$\Sigma_{\rm SFR} = A (\Sigma_{\rm gas})^N$, which in log space translates 
into a simple linear relation. The coefficient $A$ traces the absolute star 
formation efficiency \citep{kennicutt1998}, and the exponent $N$ relates the 
star formation rate to the gas density present and tells us how close the 
two variables are to linearity. We use the least-square bisector method, 
which is suitable for two independent variables \citep{isobe1990}, for the fit 
in log space and leave $\log A$ and $N$ as free parameters. It is important 
to note that for the fitting procedure alone (not the plotting), we rescaled
$\Sigma_{\rm gas}$ by a factor of 10, such that $\Sigma_{\rm gas}/10~\rm 
M_{\sun}~pc^{-2}$. We made this rescaling to minimize the covariance of 
$\log A$ and $N$ in the fit\footnote{This rescaling has no influence on the 
results of the exponent $N$, so only the coefficient $A$ will need to be 
corrected by the factor of 10.} and to make our results comparable to those of 
\citet{bigiel2008} for M63. They used the same FUV and H\,I data as we did, but 
the CO data were taken from the HERACLES CO(2--1) database \citep{leroy2009} 
with a $11\arcsec$ resolution and a CO(2--1)/CO(1--0) line ratio of 0.8. 
Our CO(1--0) data reach much deeper sensitivity than the HERACLES CO(2--1) 
data, which is essential for properly characterizing the K-S relations over all 
the optical disk and especially beyond the $r_{25}$ limit.

%

%

The best-fitting bisector linear relations, $\log \Sigma_{\rm SFR} = \log A + 
N \log \Sigma_{\rm gas}$, obtained independently for the K-S relation over the 
full radial cut out to $r_{25}$ and the UV region in the outer disk of M63 are 
shown in Fig.~\ref{fig:K-S_laws}. The errors we provide on $\log A$ and $N$ are 
$1\,\sigma$ quotes of the nearly Gaussian distributions of $\log A$ and $N$ 
obtained through Monte-Carlo simulations of 5000 data sets created from the 
original data, with random values generated within their error bars. The
scatter in our fits typically are on the order of 0.3~dex. The corresponding 
K-S relation results are presented in Table~\ref{tab:fitting-results} for both 
$\Sigma_{\rm gas} = \Sigma_{\rm H_2}$ and $\Sigma_{\rm gas } = 
\Sigma_{\rm H\,I+H_2}$, and are compared to the respective K-S relations found 
by \citet{bigiel2008} over the radial cut of the M63 disk out to $r_{\rm gal} = 
0.68\,r_{25}$ only. The K-S relations obtained over the radial cut have very 
comparable slopes (in log space) $N\sim 1.0$ for $\Sigma_{\rm H_2}$ and 
$N\sim 1.6$ for $\Sigma_{\rm H\,I+H_2}$, with our slopes systematically 
slightly steeper. They also have comparable coefficients $\log A \sim -2.0$ for 
$\Sigma_{\rm H_2}$ and $\log A \sim -2.5$ for $\Sigma_{\rm H\,I+H_2}$, with our 
coefficients systematically smaller by about $0.2-0.3~\rm dex$, but still 
within the fit scatter. The one-to-one linearity of the K-S relation between 
the SFR surface density and the molecular gas surface density is really solid, 
meaning that the gas is being consumed at a nearly constant rate, as expected 
for $\Sigma_{\rm H_2}$.

When comparing these results to the K-S relations of the UV region in the outer 
disk of M63 at the galactocentric radius $r_{\rm gal} = 1.36\,r_{25}$, we note 
that they are significantly different. The slope $N = 4.7$ (in log space) of 
the K-S relation for $\Sigma_{\rm H\,I+H_2}$ is considerably steeper in the 
external UV region. A similar trend is observed for $\Sigma_{\rm H_2}$. It is 
true that this relies on two data points, but if we trust our $3\,\sigma$ upper 
limits on $\Sigma_{\rm H_2}$, we can hardly expect to reconcile these external 
UV region measurements with the K-S relations observed over the radial cut, 
more especially as the derived star formation rate surface densities are very 
reliable. It is very likely that the CO-to-H$_2$ conversion factor adopted over 
the external UV region is underestimated, given the low metallicity expected in 
the outskirts of M63 as supported by the metallicity gradient inferred in 
M63 (see Sect.~\ref{sect:M63}) and the observed trend for an increase in 
$X_{\rm CO}$ with decreasing metallicity \citep{sandstrom2013,bolatto2013}. 
While this may induce a shift in the external UV region data points toward 
higher gas surface densities, the slopes of the K-S relations for 
$\Sigma_{\rm H\,I+H_2}$ and $\Sigma_{\rm H_2}$ will remain unchanged, unless 
there is a significant $X_{\rm CO}$ gradient over the external UV region. As a 
result, a broken K-S power law between the inner and outer regions of the M63 
disk seems to robustly emerge. It suggests that the regime of star formation 
drastically changes beyond the isophotal radius, $r_{25}$. Indeed, the 
exponents $N$ observed for the K-S relations in the external UV region show a 
nonlinear SFR regime, in clear contrast to the quasi-linear SFR regime in the 
inner regions of the disk. What triggers the 
``quenching'' of star formation in the outer regions of the M63 disk and,
consequently, the severe deviation from the commonly accepted star formation 
relation as traced by the K-S relations observed along the radial cut of M63? 
We can invoke the flaring of the outer gas layers or, more speculative, the 
possible presence of high turbulence providing support against gravitational 
collapse as suggested by \citet{longmore2013}.

To better appreciate the change in the star formation regime between the inner 
and outer regions of the M63 disk, in Fig.~\ref{fig:K-S_laws} we also plot 
``isochrones'' of constant star formation efficiencies, indicating the levels 
of $\Sigma_{\rm SFR}$ needed to consume 100\%, 10\%, and 1\% of the total 
amount of gas (corrected by a factor of 1.36 to account for helium) within 
$10^8$ years. These isochrones can also be interpreted as constant gas 
depletion timescales (i.e., the time needed to consume the total amount of gas) 
of $10^8$~yr, $10^9$~yr, and $10^{10}$ yr from top to bottom. We can see that 
when the total gas is considered (H\,I + H$_2$), the SFE settles the gas 
consumption within $10^8$ years to 1\% to 10\% for the radial cut of M63, 
whereas the external UV region has a much lower SFE with less than 1\% of the 
gas converted into stars within $10^8$ years. The drop in the star formation 
efficiency beyond the isophotal radius, $r_{25}$, can also be appreciated in 
Fig.~\ref{fig:SFE-H2HI} (upper panel). The SFE could even be lower in the 
external UV region, since the CO-to-H$_2$ conversion factor may, in reality, be 
up to 100 times more (in the extreme case) than the ``Galactic'' value 
\citep{bolatto2013} and so are the computed $\Sigma_{\rm H_2}$ values, 
depending on the metallicity of the external UV region, which may well be as 
low as 10\% to 20\% of the local Galactic ISM\footnote{When extrapolating the 
metallicity gradient determined for M63 (see Sect.~\ref{sect:M63}), we get a 
metallicity of 20\% of the local Galactic ISM at $r_{\rm gal} = 1.36\,r_{25}$, 
the galactocentric radius of the bright UV region.}. Consequently, the star 
formation still occurs in the outer regions of the M63 disk, but at very low 
efficiency, significantly lower than in the inner regions of the disk. 

%

\section{Summary and conclusions}
\label{sect:conclusion}

Deep CO(1--0) and CO(2--1) observations obtained on the IRAM 30~m telescope of 
the M63 spiral galaxy characterized by an XUV disk extending out to 2.5 times 
the optical equivalent radius were presented. We performed both a CO mapping 
along the major axis of the M63 disk from the center out to the galactocentric 
radius $r_{\rm gal} = 572\arcsec = 1.6\,r_{25}$ and over a bright UV region in 
the outer disk of M63 at the galactocentric radius $r_{\rm gal} = 
1.36\,r_{25}$. Our objective was to search for CO emission and hence for 
molecular gas in the outer regions of the M63 disk beyond the optical radius, 
where evidence of star formation is brought by both the extended UV emission 
and high H\,I column densities observed in these regions. We highlighted the 
importance of a CO detection in regions far away from the center of the galaxy, 
where the metallicity, gas temperature, excitation, and gas density are 
supposed to be lower and where stars have more difficulty to form. To 
complement our CO observations, we used FUV, NUV, H$\alpha$, 24~$\mu$m, and 
H\,I data from the literature. This allowed us to investigate the 
Kennicutt-Schmidt relations across the galaxy and beyond the isophotal radius, 
$r_{25}$, in the bright UV region. Our main results are as follows.

\begin{enumerate}
\item The CO(1--0) emission is clearly detected along the major axis of the M63 
disk out to the isophotal radius, $r_{25}$, but not beyond. However, the 
CO(1--0) is again detected in the bright UV region in the outer disk of M63 at
$r_{\rm gal} = 1.36\,r_{25}$. This is the fourth molecular gas detection in the 
outskirts of nearby spiral galaxies. The CO(2--1) emission is, on the other 
hand, confined to $r_{\rm gal} = 0.68\,r_{25}$ and thus suggests subthermal 
excitation in the outer regions of the M63 disk.

\item The radial profiles of the CO emission and other star formation tracers 
(Fig.~\ref{fig:radial-profiles}) show a severe drop with the galactocentric 
radius, in contrast to the atomic gas. Close to the $r_{25}$ limit, CO and 
the star formation tracers begin to vanish considerably, and beyond $r_{25}$, 
they all are practically absent with the exception of the faint UV emission 
and H\,I. The UV region at $r_{\rm gal} =  1.36\,r_{25}$, in which the CO 
emission is detected, is characterized by FUV and NUV emission fluxes similar 
to the fluxes observed at $r_{25}$, but stronger than the fluxes observed at 
$r_{\rm gal} > r_{25}$ along the M63 major axis. This probably reflects a tight 
correlation between the CO and UV fluxes, namely between the intensity of star 
formation and the amount of molecular gas, so it strongly suggests that the 
absence of CO detection at $r_{\rm gal} > r_{25}$, where the XUV is weaker, is 
simply the result of the CO detection threshold that is still too high. 

\item The external UV region is characterized by a very high H\,I flux with 
respect to the measured CO flux. This leads us to speculate that H\,I is more 
likely the precursor of H$_2$ rather than the product of UV photodissociation, 
since it seems to dominate in quantity. This is, however, true as long as 
substantial H$_2$ is not hidden in the outer disk regions of M63, which may be 
the case as we observe hints for an excitation temperature decrease at large 
galactocentric radii, which may lead to very weak CO lines.

\item With the integrated CO line flux measurements and the complementary 
data from the literature, we derive SFR, H\,I, and H$_2$ surface densities all 
along the major axis of the M63 disk and in the external UV region. We observe 
that the gas surface density along the radial cut is dominated by the molecular 
gas, whereas in the UV region it is dominated by the atomic gas. The 
best-fit Kennicutt-Schmidt relations, $\Sigma_{\rm SFR} = 
A (\Sigma_{\rm gas})^N$, show a broken power law from the inner to the outer 
regions of the M63 disk (Fig.~\ref{fig:K-S_laws}). Indeed, the almost linear 
K-S relation (with a slope of nearly 1 in log space) observed over the radial 
cut, in the inner regions of the disk characterized by high gas densities, 
cannot be extrapolated to the outer disk regions. The latter are 
characterized by a nonlinear SFR regime (with a K-S slope much higher than 1 in 
log space), perhaps owing to the flaring of the outer gas layers. This is the 
first time that the K-S relation is quantified in the outskirts of a spiral 
galaxy, i.e., in low gas density environments. At a molecular gas surface 
density as low as $\Sigma_{\rm H_2} = 0.35~\rm M_{\sun}~pc^{-2}$, well below 
all the determined H$_2$ surface densities referenced in spiral galaxies so far 
\citep{bigiel2008,bigiel2011}, star formation still occurs spontaneously.

\item The change in the star formation regime between the inner and outer 
regions of the M63 disk can also be appreciated by the difference in their star 
formation efficiencies. Indeed, along the major axis of the disk out to the 
isophotal radius the SFE settles the gas consumption within $10^8$ years from 
1\% to 10\%, whereas in the external UV region much less than 1\% of the gas 
is converted into stars within $10^8$ years. Consequently, star formation still 
occurs in the outer regions of the disk, but at very low efficiency.
\end{enumerate}

%

\begin{acknowledgements}
C.V.\ wishes to acknowledge support from CNRS and CONICYT through an agreement 
signed on December 11, 2007. We warmly thank the IRAM 30~m telescope staff for 
their support during the observations. We thank the anonymous referee for 
her/his very careful and constructive report. 
\end{acknowledgements}
%

\bibliographystyle{aa}
\bibliography{M63-paper-references}

\begin{thebibliography}{47}
\expandafter\ifx\csname natexlab\endcsname\relax\def\natexlab#1{#1}\fi

\bibitem[{{Alberts} {et~al.}(2011){Alberts}, {Calzetti}, {Dong}, {Johnson},
  {Dale}, {Bianchi}, {Chandar}, {Kennicutt}, {Meurer}, {Regan}, \&
  {Thilker}}]{alberts2011}
{Alberts}, S., {Calzetti}, D., {Dong}, H., {et~al.} 2011, \apj, 731, 28

\bibitem[{{Allen}(1996)}]{allen1996}
{Allen}, R.~J. 1996, {Cold dust and Galaxy Morphology} (D. L. Block \& J. M.
  Greenberg, Astrophysics and Space Science Library 209, p.50)

\bibitem[{{Allen} {et~al.}(1986){Allen}, {Atherton}, \& {Tilanus}}]{allen1986}
{Allen}, R.~J., {Atherton}, P.~D., \& {Tilanus}, R.~P.~J. 1986, \nat, 319, 296

\bibitem[{{Allen} {et~al.}(2004){Allen}, {Heaton}, \& {Kaufman}}]{allen2004}
{Allen}, R.~J., {Heaton}, H.~I., \& {Kaufman}, M.~J. 2004, \apj, 608, 314

\bibitem[{{Battaglia} {et~al.}(2006){Battaglia}, {Fraternali}, {Oosterloo}, \&
  {Sancisi}}]{battaglia2006}
{Battaglia}, G., {Fraternali}, F., {Oosterloo}, T., \& {Sancisi}, R. 2006,
  \aap, 447, 49

\bibitem[{{Bigiel} {et~al.}(2010){Bigiel}, {Leroy}, {Walter}, {Blitz},
  {Brinks}, {de Blok}, \& {Madore}}]{bigiel2010}
{Bigiel}, F., {Leroy}, A., {Walter}, F., {et~al.} 2010, \aj, 140, 1194

\bibitem[{{Bigiel} {et~al.}(2008){Bigiel}, {Leroy}, {Walter}, {Brinks}, {de
  Blok}, {Madore}, \& {Thornley}}]{bigiel2008}
{Bigiel}, F., {Leroy}, A., {Walter}, F., {et~al.} 2008, \aj, 136, 2846

\bibitem[{{Bigiel} {et~al.}(2011){Bigiel}, {Leroy}, {Walter}, {Brinks}, {de
  Blok}, {Kramer}, {Rix}, {Schruba}, {Schuster}, {Usero}, \&
  {Wiesemeyer}}]{bigiel2011}
{Bigiel}, F., {Leroy}, A.~K., {Walter}, F., {et~al.} 2011, \apjl, 730, L13

\bibitem[{{Bolatto} {et~al.}(2013){Bolatto}, {Wolfire}, \&
  {Leroy}}]{bolatto2013}
{Bolatto}, A.~D., {Wolfire}, M., \& {Leroy}, A.~K. 2013, \araa, 51, 207

\bibitem[{{Braine} {et~al.}(2007){Braine}, {Ferguson}, {Bertoldi}, \&
  {Wilson}}]{braine2007}
{Braine}, J., {Ferguson}, A.~M.~N., {Bertoldi}, F., \& {Wilson}, C.~D. 2007,
  \apjl, 669, L73

\bibitem[{{Braine} {et~al.}(2010){Braine}, {Gratier}, {Kramer}, {Schuster},
  {Tabatabaei}, \& {Gardan}}]{braine2010}
{Braine}, J., {Gratier}, P., {Kramer}, C., {et~al.} 2010, \aap, 520, A107

\bibitem[{{Braine} \& {Herpin}(2004)}]{braine2004}
{Braine}, J. \& {Herpin}, F. 2004, \nat, 432, 369

\bibitem[{{Combes} \& {Pineau Des Forets}(2000)}]{combes2000}
{Combes}, F. \& {Pineau Des Forets}, G., eds. 2000, {Molecular Hydrogen in
  Space}

\bibitem[{{Crosthwaite} {et~al.}(2002){Crosthwaite}, {Turner}, {Buchholz},
  {Ho}, \& {Martin}}]{crosthwaite2002}
{Crosthwaite}, L.~P., {Turner}, J.~L., {Buchholz}, L., {Ho}, P.~T.~P., \&
  {Martin}, R.~N. 2002, \aj, 123, 1892

\bibitem[{{Cuillandre} {et~al.}(2001){Cuillandre}, {Lequeux}, {Allen},
  {Mellier}, \& {Bertin}}]{cuillandre2001}
{Cuillandre}, J.-C., {Lequeux}, J., {Allen}, R.~J., {Mellier}, Y., \& {Bertin},
  E. 2001, \apj, 554, 190

\bibitem[{{Dale} {et~al.}(2009){Dale}, {Cohen}, {Johnson}, {Schuster},
  {Calzetti}, {Engelbracht}, {Gil de Paz}, {Kennicutt}, {Lee}, {Begum},
  {Block}, {Dalcanton}, {Funes}, {Gordon}, {Johnson}, {Marble}, {Sakai},
  {Skillman}, {van Zee}, {Walter}, {Weisz}, {Williams}, {Wu}, \&
  {Wu}}]{dale2009}
{Dale}, D.~A., {Cohen}, S.~A., {Johnson}, L.~C., {et~al.} 2009, \apj, 703, 517

\bibitem[{{de Blok} \& {Walter}(2003)}]{deblok2003}
{de Blok}, W.~J.~G. \& {Walter}, F. 2003, \mnras, 341, L39

\bibitem[{{Dickman} {et~al.}(1986){Dickman}, {Snell}, \&
  {Schloerb}}]{dickman1986}
{Dickman}, R.~L., {Snell}, R.~L., \& {Schloerb}, F.~P. 1986, \apj, 309, 326

\bibitem[{{Dong} {et~al.}(2008){Dong}, {Calzetti}, {Regan}, {Thilker},
  {Bianchi}, {Meurer}, \& {Walter}}]{dong2008}
{Dong}, H., {Calzetti}, D., {Regan}, M., {et~al.} 2008, \aj, 136, 479

\bibitem[{{Ferguson} {et~al.}(1998){Ferguson}, {Wyse}, {Gallagher}, \&
  {Hunter}}]{ferguson1998}
{Ferguson}, A.~M.~N., {Wyse}, R.~F.~G., {Gallagher}, J.~S., \& {Hunter}, D.~A.
  1998, \apjl, 506, L19

\bibitem[{{Gil de Paz} {et~al.}(2007){Gil de Paz}, {Boissier}, {Madore},
  {Seibert}, {Joe}, {Boselli}, {Wyder}, {Thilker}, {Bianchi}, {Rey}, {Rich},
  {Barlow}, {Conrow}, {Forster}, {Friedman}, {Martin}, {Morrissey}, {Neff},
  {Schiminovich}, {Small}, {Donas}, {Heckman}, {Lee}, {Milliard}, {Szalay}, \&
  {Yi}}]{gildepaz2007}
{Gil de Paz}, A., {Boissier}, S., {Madore}, B.~F., {et~al.} 2007, \apjs, 173,
  185

\bibitem[{{Gil de Paz} {et~al.}(2005){Gil de Paz}, {Madore}, {Boissier},
  {Swaters}, {Popescu}, {Tuffs}, {Sheth}, {Kennicutt}, {Bianchi}, {Thilker}, \&
  {Martin}}]{gildepaz2005}
{Gil de Paz}, A., {Madore}, B.~F., {Boissier}, S., {et~al.} 2005, \apjl, 627,
  L29

\bibitem[{{Helfer} {et~al.}(2003){Helfer}, {Thornley}, {Regan}, {Wong},
  {Sheth}, {Vogel}, {Blitz}, \& {Bock}}]{helfer2003}
{Helfer}, T.~T., {Thornley}, M.~D., {Regan}, M.~W., {et~al.} 2003, \apjs, 145,
  259

\bibitem[{{Henry} \& {Worthey}(1999)}]{henry1999}
{Henry}, R.~B.~C. \& {Worthey}, G. 1999, \pasp, 111, 919

\bibitem[{{Isobe} {et~al.}(1990){Isobe}, {Feigelson}, {Akritas}, \&
  {Babu}}]{isobe1990}
{Isobe}, T., {Feigelson}, E.~D., {Akritas}, M.~G., \& {Babu}, G.~J. 1990, \apj,
  364, 104

\bibitem[{{Kennicutt}(1989)}]{kennicutt1989}
{Kennicutt}, Jr., R.~C. 1989, \apj, 344, 685

\bibitem[{{Kennicutt}(1998)}]{kennicutt1998}
{Kennicutt}, Jr., R.~C. 1998, \apj, 498, 541

\bibitem[{{Kennicutt} {et~al.}(2008){Kennicutt}, {Lee}, {Funes}, {Sakai}, \&
  {Akiyama}}]{kennicutt2008}
{Kennicutt}, Jr., R.~C., {Lee}, J.~C., {Funes}, Jos{\'e}~G., S.~J., {Sakai},
  S., \& {Akiyama}, S. 2008, \apjs, 178, 247

\bibitem[{{Leroy} {et~al.}(2009){Leroy}, {Walter}, {Bigiel}, {Usero}, {Weiss},
  {Brinks}, {de Blok}, {Kennicutt}, {Schuster}, {Kramer}, {Wiesemeyer}, \&
  {Roussel}}]{leroy2009}
{Leroy}, A.~K., {Walter}, F., {Bigiel}, F., {et~al.} 2009, \aj, 137, 4670

\bibitem[{{Leroy} {et~al.}(2008){Leroy}, {Walter}, {Brinks}, {Bigiel}, {de
  Blok}, {Madore}, \& {Thornley}}]{leroy2008}
{Leroy}, A.~K., {Walter}, F., {Brinks}, E., {et~al.} 2008, \aj, 136, 2782

\bibitem[{{Longmore} {et~al.}(2013){Longmore}, {Bally}, {Testi}, {Purcell},
  {Walsh}, {Bressert}, {Pestalozzi}, {Molinari}, {Ott}, {Cortese}, {Battersby},
  {Murray}, {Lee}, {Kruijssen}, {Schisano}, \& {Elia}}]{longmore2013}
{Longmore}, S.~N., {Bally}, J., {Testi}, L., {et~al.} 2013, \mnras, 429, 987

\bibitem[{{Lupton} {et~al.}(2004){Lupton}, {Blanton}, {Fekete}, {Hogg},
  {O'Mullane}, {Szalay}, \& {Wherry}}]{lupton2004}
{Lupton}, R., {Blanton}, M.~R., {Fekete}, G., {et~al.} 2004, \pasp, 116, 133

\bibitem[{{Martin} \& {Kennicutt}(2001)}]{martin2001}
{Martin}, C.~L. \& {Kennicutt}, Jr., R.~C. 2001, \apj, 555, 301

\bibitem[{{Moustakas} {et~al.}(2010){Moustakas}, {Kennicutt}, {Tremonti},
  {Dale}, {Smith}, \& {Calzetti}}]{moustakas2010}
{Moustakas}, J., {Kennicutt}, Jr., R.~C., {Tremonti}, C.~A., {et~al.} 2010,
  \apjs, 190, 233

\bibitem[{{Nieten} {et~al.}(2006){Nieten}, {Neininger}, {Gu{\'e}lin},
  {Ungerechts}, {Lucas}, {Berkhuijsen}, {Beck}, \& {Wielebinski}}]{nieten2006}
{Nieten}, C., {Neininger}, N., {Gu{\'e}lin}, M., {et~al.} 2006, \aap, 453, 459

\bibitem[{{Pilyugin} \& {Thuan}(2005)}]{pilyugin2005}
{Pilyugin}, L.~S. \& {Thuan}, T.~X. 2005, \apj, 631, 231

\bibitem[{{Sancisi} {et~al.}(2008){Sancisi}, {Fraternali}, {Oosterloo}, \& {van
  der Hulst}}]{sancisi2008}
{Sancisi}, R., {Fraternali}, F., {Oosterloo}, T., \& {van der Hulst}, T. 2008,
  \aapr, 15, 189

\bibitem[{{Sandstrom} {et~al.}(2013){Sandstrom}, {Leroy}, {Walter}, {Bolatto},
  {Croxall}, {Draine}, {Wilson}, {Wolfire}, {Calzetti}, {Kennicutt}, {Aniano},
  {Donovan Meyer}, {Usero}, {Bigiel}, {Brinks}, {de Blok}, {Crocker}, {Dale},
  {Engelbracht}, {Galametz}, {Groves}, {Hunt}, {Koda}, {Kreckel}, {Linz},
  {Meidt}, {Pellegrini}, {Rix}, {Roussel}, {Schinnerer}, {Schruba}, {Schuster},
  {Skibba}, {van der Laan}, {Appleton}, {Armus}, {Brandl}, {Gordon}, {Hinz},
  {Krause}, {Montiel}, {Sauvage}, {Schmiedeke}, {Smith}, \&
  {Vigroux}}]{sandstrom2013}
{Sandstrom}, K.~M., {Leroy}, A.~K., {Walter}, F., {et~al.} 2013, \apj, 777, 5

\bibitem[{{Schruba} {et~al.}(2011){Schruba}, {Leroy}, {Walter}, {Bigiel},
  {Brinks}, {de Blok}, {Dumas}, {Kramer}, {Rosolowsky}, {Sandstrom},
  {Schuster}, {Usero}, {Weiss}, \& {Wiesemeyer}}]{schruba2011}
{Schruba}, A., {Leroy}, A.~K., {Walter}, F., {et~al.} 2011, \aj, 142, 37

\bibitem[{{Smith} {et~al.}(2000){Smith}, {Allen}, {Bohlin}, {Nicholson}, \&
  {Stecher}}]{smith2000}
{Smith}, D.~A., {Allen}, R.~J., {Bohlin}, R.~C., {Nicholson}, N., \& {Stecher},
  T.~P. 2000, \apj, 538, 608

\bibitem[{{Solomon} {et~al.}(1997){Solomon}, {Downes}, {Radford}, \&
  {Barrett}}]{solomon1997}
{Solomon}, P.~M., {Downes}, D., {Radford}, S.~J.~E., \& {Barrett}, J.~W. 1997,
  \apj, 478, 144

\bibitem[{{Solomon} {et~al.}(1987){Solomon}, {Rivolo}, {Barrett}, \&
  {Yahil}}]{solomon1987}
{Solomon}, P.~M., {Rivolo}, A.~R., {Barrett}, J., \& {Yahil}, A. 1987, \apj,
  319, 730

\bibitem[{{Thilker} {et~al.}(2005){Thilker}, {Bianchi}, {Boissier}, {Gil de
  Paz}, {Madore}, {Martin}, {Meurer}, {Neff}, {Rich}, {Schiminovich},
  {Seibert}, {Wyder}, {Barlow}, {Byun}, {Donas}, {Forster}, {Friedman},
  {Heckman}, {Jelinsky}, {Lee}, {Malina}, {Milliard}, {Morrissey}, {Siegmund},
  {Small}, {Szalay}, \& {Welsh}}]{thilker2005}
{Thilker}, D.~A., {Bianchi}, L., {Boissier}, S., {et~al.} 2005, \apjl, 619, L79

\bibitem[{{Walter} {et~al.}(2008){Walter}, {Brinks}, {de Blok}, {Bigiel},
  {Kennicutt}, {Thornley}, \& {Leroy}}]{walter2008}
{Walter}, F., {Brinks}, E., {de Blok}, W.~J.~G., {et~al.} 2008, \aj, 136, 2563

\bibitem[{{Young} \& {Scoville}(1982)}]{young1982}
{Young}, J.~S. \& {Scoville}, N. 1982, \apj, 258, 467

\bibitem[{{Young} \& {Scoville}(1991)}]{young1991}
{Young}, J.~S. \& {Scoville}, N.~Z. 1991, \araa, 29, 581

\bibitem[{{Young} {et~al.}(1995){Young}, {Xie}, {Tacconi}, {Knezek}, {Viscuso},
  {Tacconi-Garman}, {Scoville}, {Schneider}, {Schloerb}, {Lord}, {Lesser},
  {Kenney}, {Huang}, {Devereux}, {Claussen}, {Case}, {Carpenter}, {Berry}, \&
  {Allen}}]{young1995}
{Young}, J.~S., {Xie}, S., {Tacconi}, L., {et~al.} 1995, \apjs, 98, 219

\end{thebibliography}

\end{document}